\documentclass[%
      aps,
      prb, %jcp,
      reprint,
      hidelinks,
      twocolumn,
      showpacs,
      ]{revtex4-2}

\newcommand{\mff}{{\mathfrak f}}

\newcommand{\tO}{t^{(0)}} 
\newcommand{\tOO}{\mathfrak{t}^{(0)}} 
\newcommand{\TO}{T^{(0)}} 
\newcommand{\mffbarprime}{\bar{\mathfrak f}'}
\newcommand{\omegabar}{\bar\omega}
\newcommand{\omegapeak}{\omega_\text{peak}}
\newcommand{\omegapeaksbe}{\omega_\text{peak}^\text{SBE}}

\newcommand{\be}{\begin{equation}}
\newcommand{\ee}{\end{equation}}
\newcommand{\bea}{\begin{eqnarray}}
\newcommand{\eea}{\end{eqnarray}}
\newcommand{\br}{{\bf r}}

\newcommand{\bj}{{\bf j}}

\newcommand{\vF}{v_\text{F}}
\newcommand{\kF}{k_\text{F}}

\usepackage{graphicx,color,amssymb,amsfonts}
\usepackage{lipsum}
\usepackage{subcaption}
\usepackage{ragged2e}
\DeclareCaptionJustification{justified}{\justifying}
\graphicspath{{./images/}}
\definecolor{red}{rgb}{1,0,0}
\definecolor{green}{rgb}{0,0.75,0}
\definecolor{pink}{rgb}{0,1,1}
\definecolor{blue}{rgb}{0,0,1}
\definecolor{orange}{rgb}{1,0.4,0}
\definecolor{maxgreen}{rgb}{0,0.75,0}
\definecolor{maxviolet}{rgb}{0.52,0,0.56}
\definecolor{maxorange}{rgb}{1,0.61,0}
\definecolor{maxred}{rgb}{0.62,0,0}
\definecolor{maxblue}{rgb}{0,0.45,0.74}
\definecolor{maxgrey}{rgb}{0.62,0.62,0.62}
\definecolor{shadecolor}{rgb}{0.72,0.72,0.72}

\definecolor{pygreen}{rgb}{0,0.5,0}
\definecolor{pyblue}{rgb}{0,0,1}
\definecolor{pywhite}{rgb}{1,1,1}
\definecolor{pyviolet}{rgb}{0.93,0.51,0.93}
\definecolor{pygrey}{rgb}{0.5,0.5,0.5}
\definecolor{pyblack}{rgb}{0,0,0}
\definecolor{pyorange}{rgb}{1,0.65,0}
\definecolor{pyyellow}{rgb}{1,1,0}
\definecolor{pyred}{rgb}{1,0,0}
\usepackage{bm}

\usepackage{amsmath}
\usepackage{amssymb}
\usepackage{amsfonts}
\usepackage{upgreek}
\usepackage{mathbbol}
\usepackage{mathtools} 
\usepackage[ colorlinks,
    linkcolor={blue},
    citecolor={blue},
    urlcolor={blue}]{hyperref}
\usepackage{bbold} 
\usepackage[normalem]{ulem}
\usepackage{caption}
\usepackage{pgf}
\usepackage{placeins}

\DeclareMathAlphabet{\mathpzc}{OT1}{pzc}{m}{it}
\DeclareMathAlphabet\mathbfcal{OMS}{cmsy}{b}{n}

\newcommand{\ci}{{ i}}

\newcommand{\bE}{{\bf E}}

\newcommand{\sdis}{\hspace{0.05em}}

\definecolor{darkred}{rgb}{0.66666667,0,0}

\newcommand\eqt{\hspace{0.17em}{=}\hspace{0.17em}}

\newcommand\proptot{\hspace{0.17em}{\propto}\hspace{0.17em}}
\newcommand\eqtexclmark{\hspace{0.17em}{\overset{!}{=}}\hspace{0.17em}}
\newcommand\cdott{\hspace{0.17em}{\cdot}\hspace{0.17em}}

\newcommand\lesssimt{\hspace{0.17em}{\lesssim}\hspace{0.17em}}
\newcommand\coloneqqt{\hspace{0.17em}{\coloneqq}\hspace{0.17em}}

\newcommand\neqt{\hspace{0.17em}{\neq}\hspace{0.17em}}

\newcommand\apt{\hspace{0.17em}{\approx}\hspace{0.17em}}

\newcommand\pmt{\hspace{0.17em}{\pm}\hspace{0.17em}}

\newcommand\pt{\hspace{0.17em}{+}\hspace{0.17em}}
\newcommand\mt{\hspace{0.17em}{-}\hspace{0.17em}}
\newcommand\llt{\hspace{0.17em}{\ll}\hspace{0.17em}}
\newcommand\letext{\hspace{0.17em}{\le}\hspace{0.17em}}
\newcommand\gt{\hspace{0.17em}{>}\hspace{0.17em}}

\newcommand\get{\hspace{0.17em}{\ge}\hspace{0.17em}}

\newcommand\intext{\hspace{0.15em}{\in}\hspace{0.15em}}

    \newcommand\timest{\hspace{0.14em}{\times}\hspace{0.09em}}
 \newcommand\rightarrowtext{\hspace{0.15em}{\rightarrow}\hspace{0.15em}}

\newcommand{\bv}{\mathbf{v}}

\newcommand{\bk}{\mathbf{k}}

\newcommand{\ggtext}{\hspace{0.17em}{\gg}\hspace{0.17em}}

\newcommand{\fchirp}{f_\text{chirp}}

\usepackage[varg]{txfonts} %font/Schrift

\DeclareUnicodeCharacter{2212}{\textendash}

\begin{document}

\title{Influence of chirp and carrier-envelope phase on non-integer high-harmonic generation
}

\author{Maximilian Graml}
\affiliation{Institute  of  Theoretical  Physics,  University  of  Regensburg, Universit\"atsstra\ss e  31,  D-93053  Regensburg,  Germany}
\author{Maximilian Nitsch}\altaffiliation{Present adress: NanoLund and Solid State Physics, Lund University, Box 118, 22100 Lund, Sweden}
\affiliation{Institute  of  Theoretical  Physics,  University  of  Regensburg, Universit\"atsstra\ss e  31,  D-93053  Regensburg,  Germany}

\author{Adrian Seith}
\author{Ferdinand Evers}
\author{Jan Wilhelm}\email{jan.wilhelm@physik.uni-regensburg.de}
\affiliation{Institute  of  Theoretical  Physics,  University  of  Regensburg, Universit\"atsstra\ss e  31,  D-93053  Regensburg,  Germany}

\date{\today}
\begin{abstract}

High harmonic generation (HHG) is a versatile technique for probing ultrafast electron dynamics. 
While HHG is sensitive to the electronic properties of the target, HHG also depends on the waveform of the laser pulse. 
As is well known, (peak) positions, $\omega$, in the high-harmonic spectrum can shift when the carrier envelope phase (CEP), $\varphi$ is varied.
We derive formul{\ae} describing the corresponding parametric dependencies of CEP shifts; in particular, we have a transparent result for the (peak) shift,
$d\omega/d\varphi\eqt{-}\sdis2\sdis \mffbarprime \omega/\omega_0$, where $\omega_0$ describes the fundamental frequency and $\mffbarprime$ characterizes the chirp of the driving laser pulse.
We compare the analytical formula to full-fledged numerical simulations finding only  17\,\% average relative absolute deviation in $d\omega/d\varphi$.
Our analytical result is fully consistent with experimental observations.

\end{abstract}

\maketitle

\section{Introduction}

\begin{figure}[b]
    \centering
    \includegraphics[width=8.6cm]{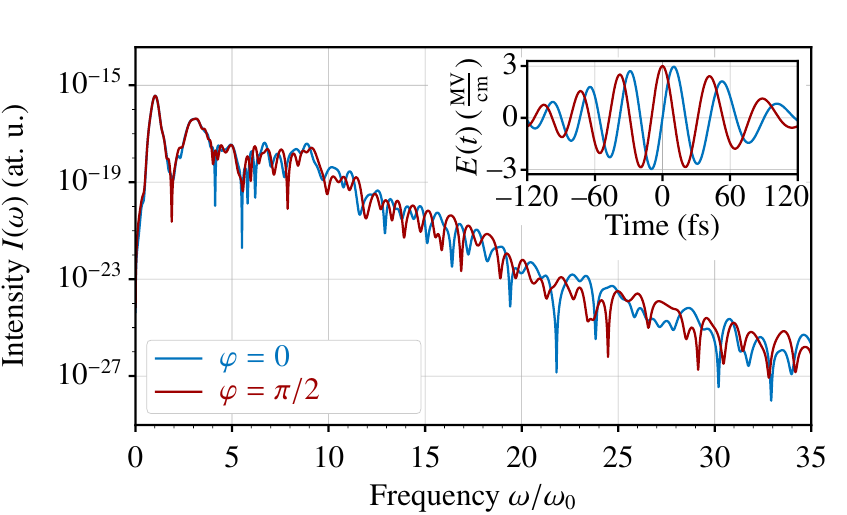}
    \caption{High-harmonic emission spectrum~$I(\omega)$  as function of the frequency~$\omega$ computed from semiconductor Bloch equations (SBE),~\cite{SchmittRink1988, Lindberg1988,Aversa1995,Schaefer2002,Haug2008,Haug2009,Kira2011,Foeldi2017,Silva2019b,Li2019,Yue2020,Thong2021,Wilhelm2021,Yue2022} Eq.~\eqref{eq-emission}.
    For the SBE simulation, we employ a two-band Hamiltonian~\cite{Schmid2021} to model the topological surface state of Bi$_2$Te$_3$.
    As driving electric field~$\bE(t)$ (inset), we use  Eq.~\eqref{eq-Efield} with $E_0\eqt 3\,$MV/cm, $\omega_0\eqt2\pi\cdott25$\,THz, $\fchirp\eqt-$1.25\,THz, $\sigma\eqt 90$\,fs as used in simulations in Ref.~\citenum{Schmid2021}.
   We employ two different CEPs, $\varphi\eqt0$ (blue) and $\varphi\eqt\pi/2$ (red). 
   The driving electric field~$E(t)$ is sketched in the inset. 
   }
     \label{f1}
\end{figure}

High harmonic generation (HHG) is a unique fingerprint of ultrafast electron dynamics in solids:~\cite{Chin2001,Ghimire2011,Schubert2014, Hohenleutner2015,Vampa2015,Luu2015,Garg2016,Yoshikawa2017,Hafez2018,Schmid2021,You2017,Sivis2017,Garg2018,Vampa2015b,Tancogne2017,Yue2022,Park2022,Luu2018,Liu2017,Silva2019,Chacon2020,Baykusheva2021a,Shirai2018,Leblanc2020,Song2019,Hollinger2020,Goulielmakis2022}
It is generated when atomically strong electric fields drive charge currents that in turn emit electromagnetic radiation. In solids, such currents are understood as interband transitions and (semiclassical) intraband currents.
 The emitted light supports frequencies much higher than those of the driving field, see also Fig.~\ref{f1} as an illustration. 
 Since high harmonics are sensing acceleration processes of the charge carriers, HHG can be used for monitoring dynamical processes. 
 The information thus incorporated allows to  reconstruct band structures;~\cite{Vampa2015b,Tancogne2017} it reflects dynamical Bloch oscillations~\cite{Schubert2014,Luu2015,Borsch2020} and  Berry phase effects.~\cite{Luu2018,Liu2017,Silva2019,Bauer2018,Drueke2019,Jurss2019,Jurss2020,Moos2020,Chacon2020,Baykusheva2021a,Baykusheva2021b,Lou2021,Bharti2022}
In the past, HHG has been analyzed to study charge carrier dynamics in dielectrics~\cite{Ghimire2011,Luu2015,You2017,Garg2018} and semiconductors.~\cite{Schubert2014, Hohenleutner2015, Vampa2015}
Fresh applications to three-dimensional topological insulators and their gapless surface states have been published recently.~\cite{Schmid2021, Bai2021}
These surface states have been argued to be an ideal platform for lightwave electronics.~\cite{Reimann2018, Schmid2021}
This is because the suppression of backscattering due to the spin-momentum locking makes it easier to facilitate quantum control for long times.~\cite{Giorgianni2016, Reimann2018,Schmid2021} 
An intriguing feature of topological surface states is the effect of the carrier-envelope phase (CEP)~\cite{Jones2000,Paulus2001,Baltuska2002,Cundiff2003,Baltuska2003,Manzoni2010,Meierhofer2022} on the high-harmonic spectrum:
upon tuning the CEP, harmonic
orders shift continuously to non-integer multiples of the driving frequency $\omega_0$.~\cite{Schmid2021}
This is illustrated in   Fig.~\ref{f1}, where we display the HHG for a topological surface state; for the two CEP-values shown, the peaks of orders 13-18 are shifted against each other by $\omega_0/2$. 
Corresponding shifts have been observed before in  semiconductors~\cite{Schubert2014,Shirai2018,Leblanc2020} and dielectrics~\cite{You2017,Garg2018,Song2019,Hollinger2020}; the particular aspect of  topological surfaces is that CEP shifts occur at relatively low harmonic order~\cite{Schmid2021}.

In this work, we develop a minimal model of high harmonic generation that explains the CEP shifts in  analytical terms. 
The main result of our work is that under a tuning of the CEP by $d\varphi$,  the frequency~$\omega$ of high harmonics shifts by
\begin{align}
\label{e0} 
   d\omega = -\,2\,\frac{\omega}{\omega_0}\,\mffbarprime\,d\varphi\,;
\end{align}
here, $\mffbarprime$ characterizes the chirp of the driving laser pulse~\cite{Zhou1996,Shin1999,Lee2001}. 
Eq.~\eqref{e0} has been derived for generic two-band models of non-interacting fermions. Remarkably, \eqref{e0} only contains parameters of the driving laser pulse indicating its applicability for a wide range of model Hamiltonians.
We show that the formula is in line with CEP shifts observed in Ref.~\citenum{Schmid2021} and with an additional, extended set of simulations. 
Thus, the assumptions underlying our minimal model are validated.
Our work thus is yet another stepping stone towards an improved understanding of the fundamental mechanisms and parametric dependencies governing HHG. 

\section{Mathematical definition of CEP shifts}

For deriving parametric dependencies of CEP shifts in high harmonics, we consider a CEP variation~$\varphi\rightarrowtext \varphi \pt d\varphi$ in the driving electric field, see inset of Fig.~\ref{f1} as an illustration.
A formal definition of  CEP shifts of high harmonics spectra $I(\omega)$ embarks on the observation that 
for a given $d\varphi$ a corresponding frequency shift $
\omega\rightarrowtext \omega \pt d\omega$ can be found that leaves the emission  unchanged, $dI\eqt0$.~\footnote{Other kinds of CEP shifts that also give useful characterizations of the $I(\omega,\varphi)$ map can be conceived, too. 
For example, rather than tracing lines with $dI\eqt 0$, one can trace maxima or minima, so requiring ~$d(\partial I/\partial\omega)_\varphi\eqt0$, see Fig.~\ref{f2}\,(b) as an example. 
In analogy to Eq.~\eqref{e12}, we then consider 
\begin{align*}
 d(\partial I/\partial\omega) =   
 (\partial^2 I/\partial\omega^2)\, d\omega + 
 (\partial^2I/(\partial\omega\partial\varphi))\, d\varphi
\end{align*}
which, together with the defining requirement $d(\partial I/\partial\omega)_\varphi\eqt0$, leads to an alternative set of lines $\omega(\varphi)$ in the $\omega$-$\varphi$ plane with tilt angle
 \begin{align*}
     \frac{d\omega}{d\varphi} \coloneqq -
     \frac{\partial^2I/(\partial\omega\partial\varphi)}{\partial^2I/\partial\omega^2}
     \,.
\end{align*}
This definition and definition~\eqref{e4} are equivalent in case maxima and minima lines are also equi-intensity lines.}
We have 
\begin{align} 
dI  = (\partial I/\partial \omega)_\varphi \,d\omega \pt (\partial I/\partial \varphi)_\omega\,d\varphi\,, 
\label{e12} 
\end{align} 
and the condition $dI\eqtexclmark 0$ translates into the definition of the frequency shift per CEP variation,
\begin{align} 
\frac{d\omega}{d\varphi}\coloneqq  - 
\frac{(\partial I/\partial \varphi)_\omega}{(\partial I/\partial \omega)_\varphi}\,. \label{e4}
\end{align} 
In general, $d\omega/d\varphi$ is a function of $\omega$ and $\varphi$; $d\omega/d\varphi$   mathematically describes the tilt angle of the  equi-intensity lines in the $(\omega,\varphi)$-plane, which is observed in CEP-dependent high-harmonic spectra; see Fig.~\ref{f2} for an illustration.

By integrating Eq.~\eqref{e4} one can find the equi-intensity line $\omega(\varphi)$ -- for a fixed initial condition of integration, e.g.~$\omegabar\coloneqqt \omega(\varphi\eqt 0)$;
we denote this by $\omega_{\omegabar}(\varphi)$. %
Intuitively speaking, $\omega_{\omegabar}(\varphi)$
is the line in the map of $I(\omega,\varphi)$ that traces the equi-intensity line crossing the point $(\bar \omega,\varphi\eqt0)$.

\section{CEP shift from SBE simulations}
We start with numerical simulations of CEP shifts in high harmonics to illustrate the phenomenon and to motivate the minimal analytical model that we introduce later.
For our theoretical analysis, we model the incoming laser pulse by  the time-dependent electric field aligned in $x$-direction 
\begin{align}
\bE(t)\eqt \hat{\mathbf{x}}\,E_0    \sin\big(\omega_0\,(1+\fchirp \sdis t\sdis)\,t + \varphi\big)\, e^{-t^2/\sigma^2}\;,
    \label{eq-Efield}
\end{align}
with the parameters field strength~$E_0$, (driving) frequency~$\omega_0$, chirp~$\fchirp$, CEP~$\varphi$, and pulse duration~$\sigma$.
We employ the two-band model for the topological surface state of Bi$_2$Te$_3$ used in Ref.~\cite{Schmid2021}; it includes a Dirac cone at the $\Gamma$-point and the hexagonal warping in the band structure of the topological surface state.~\cite{Liu2010}
Taking the pulse form and the model Hamiltonian as an input, we solve the semiconductor Bloch equations (SBE),~\cite{SchmittRink1988, Lindberg1988,Aversa1995,Schaefer2002,Haug2008,Haug2009,Kira2011,Foeldi2017,Li2019,Silva2019b,Yue2020,Thong2021,Wilhelm2021,Yue2022} yielding the time-dependent density matrix~$\rho(t)$. From this we obtain  
the physical current density~$\bj(t)$ and the emission spectrum~$I(\omega)$,~\cite{Wilhelm2021}
\begin{align}
   \bj(t)\coloneqq \frac{-e}{V}\, \text{Tr}\sdis(\sdis\rho(t)\sdis\dot\br)\;,
   \hspace{2em} 
   I(\omega)= \frac{\omega^2}{3c^2}\,|\sdis\bj(\omega)|^2\,,
   \label{eq-emission}
\end{align}
where $-e/V$ is the electron charge density, $\dot\br$ the velocity operator,  $c$ the speed of light and $\bj(\omega)$  the Fourier transform of $\bj(t)$. 
We checked the convergence of observables with numerical parameters, see App.~\ref{app:convergencetests}.

\begin{figure}
    \centering
    \includegraphics[width=8.6cm]{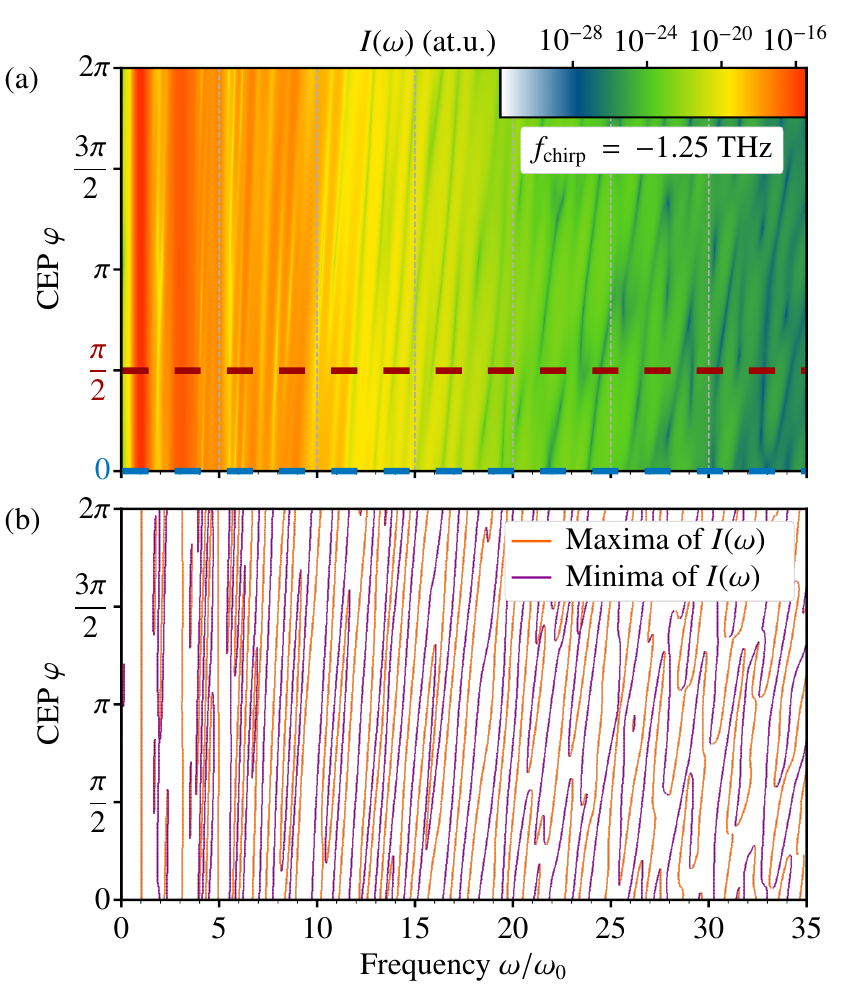}
    \caption{   (a) High-harmonics spectrum $I(\omega)$ computed from SBE, Eq.~\eqref{eq-emission} for 384 discrete CEPs~$\varphi\intext[0,2\pi]$ with $\bE(t)$ and parameters as in Fig.~\ref{f1}. 
    The heat map along the blue and red horizontal line (CEP~$\varphi\eqt0$ and $\varphi\eqt\pi/2$) represents the emission spectra from Fig.~\ref{f1}.
 (b) Local extrema of $I(\omega)$ from (a).}
    \label{f2}
\end{figure}
The resulting high-harmonics spectrum~$I(\omega)$ for  pulse parameters adapted to experiment~\cite{Schmid2021} is shown in Fig.\,\ref{f1}, for a sine-like pulse ($\varphi\eqt0$) and a cosine-like pulse ($\varphi\eqt\pi/2$):
Both high-harmonics spectra are similar up to fifth harmonic order, $\nu \eqt5$, using a dimensionless frequency $\nu\coloneqqt\omega/\omega_0$. 
At higher frequencies, $14\lesssimt {\nu} \lesssimt20$, the two spectra differ in the sense that the maximum of one coincides with the minimum of the other.  
At even higher frequencies,  $30\lesssimt{\nu }\lesssimt 35$, maxima of the two spectra coincide and minima also coincide.

Similar to the experiment,~\cite{Schmid2021} we continuously vary the CEP from 0 to $2\pi$, see Fig.~\ref{f2}\,(a) and~(b).  
We confirm the main experimental findings, albeit here observed in a much larger window, $5\lesssimt\nu\lesssimt35$, instead of $12\lesssimt\nu\lesssimt21$ in Ref.~\citenum{Schmid2021}:  
The frequency shift grows at increasing harmonic order, which eventually leads to a pattern of tilted lines with tilt angle growing from left to right in Fig.~\ref{f2}\,(b). Indications of an increase of the tilt-angle have been observed before in semiconductors and dielectrics, but the patterns there are less pronounced and systematical.~\cite{You2017} 
Presumably this is why a systematic theoretical understanding predicting parametric dependencies of CEP shifts has not been worked out.

\section{CEP shifts for a semiclassical model -- analytical formula}\label{secIV}

The systematic growth of the tilt angle with the high-harmonic order seen in Fig.~\ref{f2}\,(b) suggests that there should be a simple analytical formula  characterizing parametric dependencies. In this section such a formula is derived within a minimal model. 
We employ a semiclassical framework~\cite{Ashcroft1976} neglecting anomalous velocity contributions.~\cite{Xiao2010}
Within this model, the electron velocity is given by 
\begin{align}
        \bv(t)  = 
    \left.
    \frac{\partial\epsilon}{\hbar\,\partial\bk}
    \right|_{\bk=\bk(t)}\,.\label{eq-current-quasiclassical}
\end{align}
$\bk(t)$ is the excursion of the electron in reciprocal space.
In semiclassics, $\bk(t)$ fully characterizes the dynamics of the electron and is given by the Bloch acceleration theorem~\cite{Bloch1929}
\begin{align}
\bk(t)= \bk_0 +\frac{1}{\hbar} \int\limits^t_{-\infty} {\bf F}(t')\,dt'\,,\label{e8}
\end{align}
where ${\bf F}$ is the acting force, ${\bf F}(t)\eqt{-}e\sdis {\bf E}(t)$, if only electric fields are to be accounted for. 
In our simplified approach we assume that the time dependence of $\bj(t)$ is captured by $\bv(t)$ taken at a characteristic wavenumber $\bk_0$. For the purpose of calculating CEP shifts, prefactors - such as effective charge densities - can be ignored since they cancel for CEP shifts in Eq.~\eqref{e4}.

We now analyze CEP shifts within the framework of model \eqref{eq-current-quasiclassical} and~\eqref{e8}.
Since it is assumed $\bj(t)\proptot\bv(t)$, 
we have 
\begin{align}
    I(\omega)\proptot \omega^2|\bv(\omega)|^2 = |\partial_t \bv(\omega)|^2   \label{e8a}
\end{align} 
from Eq.~\eqref{eq-emission}, where 
\begin{align}
\partial_t \bv(\omega) \coloneqq \int dt\ e^{-\ci \omega t}\ \partial_t \bv(t) = \int dt\ e^{-\ci \omega t} \ \partial_t 
\hspace{-0.15em}\left. \frac{\partial \epsilon}{\hbar\, \partial \bk}\right|_{\bk(t)}
\end{align}
is the Fourier transform of the time-dependent acceleration. 

If the $\bk$-derivative is analytic in the range of excursion of $\bk(t)$, we may simplify 
\begin{align*}
    \partial_t \sdis v_i\sdis(\omega) &= \sum_{j} 
    \int dt\ e^{-\ci\omega t}
  \left.  \frac{\partial^2\epsilon}{\hbar\,\partial k_i\sdis \partial k_{\hspace{-0.07em}j}}
    \right|_{\bk=\bk(t)}
    \partial_t k_{\hspace{-0.07em}j}\sdis(t) \nonumber\\
    &=  -\frac{e}{\hbar} \sum_{j}
    \int dt\ e^{-\ci\omega t} \left.\frac{\partial^2\epsilon}{\hbar\,\partial k_i\sdis \partial k_{\hspace{-0.07em}j}} 
     \right|_{\bk=\bk(t)}
    E_{\hspace{-0.1em}j}\sdis(t). 
\end{align*}
where we have  $\partial_tk_{\hspace{-0.07em}j}\sdis(t)\eqt {-}\sdis e E_{\hspace{-0.1em}j}\sdis(t)/\hbar$ in the absence of magnetic fields. It is a necessary condition for the generation of high harmonics that $\partial^2\epsilon/(\partial k_i \partial k_j)    |_{\bk=\bk(t)}$ is time dependent. 
For the special case of parabolic dispersions with isotropic effective mass $m$ we obtain
\begin{align*}
    \partial_t \bv(\omega) 
    &=  -\frac{e}{m}\, \bE(\omega)\,.
\end{align*}
The conclusion is that parabolic dispersions do not exhibit HHG (within the validity of our minimal model)~\cite{Ghimire2012}. 

As a minimal model for Dirac fermions we consider a linear dispersion
\begin{align}
\epsilon(\bk)=\hbar\vF\sdis|\bk|\,.\label{eq-lindisp}
\end{align}
The velocity~\eqref{eq-current-quasiclassical} for the linear dispersion 
is
\begin{align}
    \bv(t)  = 
    \hat{\mathbf{x}}\,\vF \, \text{sgn}(k_x(t))\label{eq-current}\,; 
\end{align}
the velocity is a constant, $\vF$, and it only changes its sign when $k_x(t)$ crosses zero.
 Hence, the acceleration $\partial_t \bv(t)$ is a sequence of $\delta$-functions in time with a corresponding Fourier transform  
\begin{align}
     \partial_t \bv(\omega) =
     \sum_{m=1}^{N_z}  \mathbf{v}_m \exp(i\omega t_m)\,,  \label{eq-current-ft}
\end{align}
where $\mathbf{v}_m\eqt 2\vF\,(-1)^{m+1}\,\hat{\mathbf{x}}$ for the linear dispersion~\eqref{eq-lindisp}. 
The summation is over the zeros
$t_m$ of $k_x(t)$, which are readily obtained from \eqref{e8}; $N_z$ denotes the number of these zeros (see App.~\ref{Appendix-Fourier-trafo} for a formal derivation.) 
The zeros $t_m$ will shift in the presence of a CEP,  $t_m(\varphi)$. Since we require a $2\pi$-periodicity in $\varphi$, we have $t_m(\pm2\pi)\eqt t_{m\pm p}(0)$, 
so after a full rotation a root $m$ shifts into root $m\pmt p$, $p$ being integer.~\footnote{Note that due to the special nature of the Dirac-dispersion, the acceleration $\partial_t \bv$ does not scale with the applied force $-e\bE$. The electric field enters only indirectly in the sense  that for non-vanishing $\bk_0$ a minimum field-strength is required to produce zeros in $\bk(t)$.}

Eq.~\eqref{eq-current-ft} is expected to be applicable to broader classes of band-structures, the main requirement is a sufficiently large frequency $\omega$. 
Indeed, high-harmonic radiation consistent with~\eqref{eq-current-ft}  has been  observed in experiments on  semiconductors~\cite{Hohenleutner2015}.
We show in detail that the following analysis also applies to the general case, Eq. \eqref{eq-current-ft} in App.~\ref{app-cepshiftsgenericbands}.

The  emission intensity~\eqref{e8a} corresponding to \eqref{eq-current-ft} for the linear dispersion~\eqref{eq-lindisp}  with $\mathbf{v}_m\eqt 2\vF\,(-1)^{m+1}\,\hat{\mathbf{x}}$ reads 
\begin{align}
I(\omega) \propto |\partial_t\bv(\omega)|^2 
     \propto \sum_{\ell,m}^{N_z} (-1)^{\ell+m}e^{i\omega\sdis t_{\ell m}(\varphi)}\label{e13}\,,
\end{align}
with $t_{\ell m}(\varphi)\coloneqqt t_\ell(\varphi)\mt t_m(\varphi)$. Then, Eq.~\eqref{e4} readily implies 
\begin{align}
\frac{d\omega}{d\varphi} = -\, \omega\, \frac{I_t^\prime}{I_t} 
\label{e16}
\end{align}
where 
\begin{align}
    I_t(\omega,\varphi)&\coloneqq \sum_{\ell,m}^{N_z}(-1)^{\ell+m} e^{i\omega t_{\ell m}(\varphi)}t_{\ell m}(\varphi)\,,\label{e17}
    \\
    I_t^\prime(\omega,\varphi) &\coloneqq
   \sum_{\ell,m}^{N_z}(-1)^{\ell+m} e^{i\omega t_{\ell m}(\varphi)}\partial_\varphi t_{\ell m}(\varphi)\,.
    \label{e18}
\end{align}

As an application, we consider the situation in which the CEP $\varphi$ induces a homogeneous shift of all roots: 
$t_m(\varphi)\eqt t_m(0) \pt \varphi/\omega_0$ and, in addition, an equidistant spacing $t_m(\pi) \eqt t_{m+1}(0)$. Here implied is that $t_{\ell m}(\varphi)$ is independent of $\varphi$ and therefore $I_t^\prime\eqt0$. 
We conclude that a non-vanishing CEP shift requires that the zeros of $\bk(t)$  are not equidistantly spaced.~\footnote{This conclusion has already been drawn in Refs.~\cite{You2017}, \cite{Shirai2018} and~\cite{Frolov2011, Frolov2012, Naumov2015,Sansone2009}.}

Non-equidistant roots of $\bk(t)$ result from a time-dependent carrier frequency, which is defined as $(1+\mff(t))\sdis\omega_0$ ("chirp" $\mff(t)$). 
In the App.~\ref{a4}, we show that for small and slowly varying~$\mff(t)$, Eq.~\eqref{e16} simplifies to 
\begin{align}
\frac{d\omega}{d\varphi} = -\sdis 2\sdis\frac{\omega}{\omega_0}\sdis \mffbarprime\,,
\label{e19}
\end{align}
where $\mffbarprime$ is the average slope of $\mff(t)$. 
Eq.~\eqref{e19} is our main result; it implies that under generic conditions the tilt angle $d\omega/d\varphi$ increases linearly in~$\omega$, and is independent of~$\varphi$. 

We now adress the shift of peak frequencies~$\omegapeak$ in the high-harmonics spectrum, when changing the CEP from $0$ to $2\pi$, $\Delta\omegapeak\coloneqqt{\int_0^{2\pi}} d\varphi \,( d\omega/d\varphi)$ along an equi-intensity line~$\omega(\varphi)$. 
In the regime $|\mffbarprime/\omega_0|\llt 1$
we focus on, $d\omega/d\varphi$ is only weakly dependent on the integration variable, because the relative change of $\omega(\varphi)$ along the equi-potential line is small: 
$\Delta\omegapeak\llt\omegapeak$; 
we thus approximate on the rhs of \eqref{e19} 
$\omega\sdis(\varphi)\apt\sdis\omega\sdis(\varphi\eqt0)\eqt\omegapeak$,
and arrive at the peak shift
\begin{align}
 \Delta \omegapeak = -\,4\pi\,\frac{\omegapeak}{\omega_0}\,\mffbarprime
\,.
    \label{e21}
\end{align}
As an application of Eqs.~\eqref{e19} and \eqref{e21}, we consider an archetypical  electric-field pulse~\eqref{eq-Efield}, implying $\mffbarprime\eqt\fchirp$,  and small filling, i.e., $\kF$ far away from all Brillouin zone boundaries, motivating a non-vanishing $|\bk_0| \lesssimt\kF$.
In this case the number of roots of $\bk(t)$, $N_z$, is bounded: at infinite times the integral in \eqref{e8} vanishes (since $\bE(\omega\eqt0)\eqt 0$) so that $\bk(t)$ takes a non-vanishing limiting value. 
 Eq.~\eqref{e19} takes a simple form; while the frequency~$\omega$ and the detuning per period $\mffbarprime/\omega_0$ enter, other material or system parameters do not. 
The absence of $\vF$ follows from the fact that the CEP is given as an intensity ratio, Eq.~\eqref{e4}. 
Similarly, the envelope parameter~$\sigma$ of the driving electric field~\eqref{eq-Efield} (parameterized via $N_z$)  cancels since our approximations imply~$I_t'\proptot I_t$. 
The linear dependency of  $d\omega$ on the chirp parameter~$\mffbarprime$ is rationalized as follows: 
By definition, the chirp accounts for the non-linear spacing of the roots of $\bk(t)$. 
Therefore, in the absence of chirp, the CEP translates all roots by the same amount and therefore can be eliminated by a redefinition of the origin of time, $t\rightarrowtext  t\mt d\varphi/\omega_0$. 
Thus, at $\mffbarprime\eqt0$ the current and the emission spectrum are both independent of~$\varphi$ resulting in the absence of CEP shifts, $d \omega\eqt 0$ (cf. App.~\ref{a4a}).
At non-vanishing chirp, corrections arise already at linear-order, $d \omega \proptot \mffbarprime$, reflecting the fact that the tilt  $d\omega/d\varphi$ can take either sign. 

Finally, the proportionality $d \omega \proptot \omega/\omega_0$ can be understood by recalling that the electrons perform a number of 
$\omega/\omega_0$ cycles during a single fundamental period of $E(t)$. They thus can be expected to be more sensitive to a parametric change in~$E(t)$, for example a change of $\mffbarprime$, by a factor of
$\omega/\omega_0$. 
\begin{figure*}[t!]
    \centering
    \includegraphics[width=17.5cm]{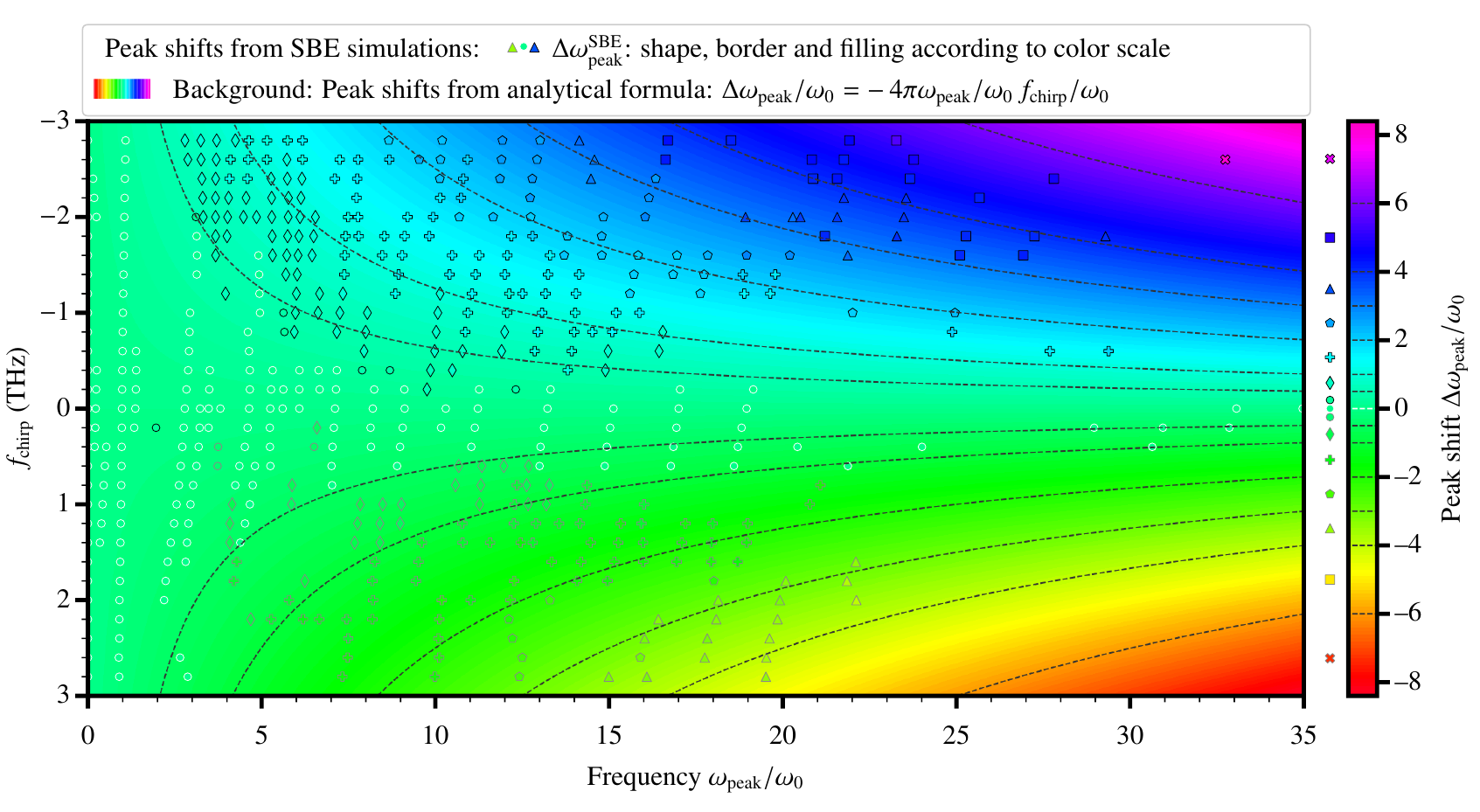}
    \caption{
   Quantitative comparison of the peak shifts as obtained from the analytical formula~\eqref{e21} (background{, $\Delta \omegapeak$}) and from SBE simulations (color-filled circles{, $\Delta \omegapeaksbe$}) in the parameter plane $(\omegapeak,\fchirp)$. 
    {$\Delta \omegapeaksbe$} are obtained by tracing continuous maximum lines of the  emission~$I(\omega)$ as shown in Fig.~\ref{f2}\,(b)  from the initial point  $(\omegapeak^\text{{SBE}},\varphi\eqt0)$ to the final point $(\omegapeak^\text{{SBE}}\pt\Delta \omegapeak^\text{{SBE}},\varphi\eqt2\pi)$. 
    The dashed lines are guiding the eye; they indicate constant $\Delta\omegapeak$ in the analytical formula. 
    (For the SBE simulations, we considered 29  chirps  with 384 CEPs each resulting in  11136 SBE runs.) 
    }
    \label{f3}
\end{figure*}

\section{comparison of the analytical formula to SBE simulations}
We proceed with a comparison of the analytical formul\ae~\eqref{e19}, \eqref{e21} to SBE simulations.
In Fig.~\ref{f2} we display SBE simulations of the CEP-dependent high-harmonic spectra.
A straight-line character of the extremal lines -- and correspondingly also the equi-intensity lines -- is seen which is synonymous with a minor tilt-angle dependency on  CEP~$\varphi$; further the tilt angle $d\omega/d\varphi$ increases with~$\omega$.
Both observations are fully consistent with our main result,  Eq.~\eqref{e19}.

With respect to the sign of the tilt, 
we further observe in Fig.~\ref{f2} that the extremal lines are tilted  to the right (from south west to north east) which implies a positive tilt angle, $d\omega/d\varphi\get 0$.
The positive sign of the tilt angle is consistent with the analytical prediction \eqref{e19}: we evaluate for the pulse \eqref{eq-Efield} the chirp $\mffbarprime\eqt\fchirp\eqt{-}\sdis1.25\,$THz implying $d\omega/d\varphi\get 0$ (for $\omega\get0$). 
The high-harmonic spectra reported in Fig.~\ref{f2} have been generated for a specific choice of the pulse parameters pulse duration~$\sigma$,  frequency~$\omega$,  chirp~$\fchirp$,  field strength~$E_0$, and  the dephasing time $T_2$~\cite{Floss2018, Wilhelm2021}.
We provide an extended set of CEP-dependent high-harmonic spectra in App.~\ref{a9}, Figs.~\ref{fig-skeletons}\,--\,\ref{fig-comparisonbite-tssbulk1d}, where we change  the pulse parameters and we also use two generic semiconductor Hamiltonians instead of the Dirac-type band structure. 
In all spectra, we observe that the tilt angles follow our analytical  results~\eqref{e19} and~\eqref{e21}.
This finding is in line with the derivation of~\eqref{e19} and~\eqref{e21}, that holds independently of a specific band structure or specific pulse shapes. 
For a quantitative comparison, we focus on peak shifts $\Delta\omegapeak$.
From the SBE simulations shown in Fig.~\ref{f2}\,(b) we extract peak shifts by tracing continuous maximum lines ("percolating lines") of the  emission~$I(\omega)$ connecting the points in the parameter plane $(\omegapeak^\text{SBE},\varphi\eqt0)$ and  $(\omegapeak^\text{SBE}\pt\Delta \omegapeak^\text{SBE},\varphi\eqt2\pi)$. We obtain pairs of $(\Delta\omegapeak^\text{SBE}/\omegapeak^\text{SBE})$ as
(1.0/10.2), (1.6/19.7), (2.2/24.7). 
Based on equation \eqref{e21} we expect a ratio
$\Delta\omegapeak/\omegapeak\eqt{-}\sdis4\pi \mffbarprime/\omega_0\eqt0.1$, in good quantitative agreement with the extracted SBE data 
(simulation parameters: $\omega_0\eqt2\pi\cdott25$\,THz and $\mffbarprime\eqt\fchirp\eqt{-}\sdis1.25$\,THz). 
We proceed and calculate the peak shifts~$\Delta\omegapeak$ for a collection of chirps to test the limits of \eqref{e21} in the plane spanned by $\omegapeak$ and $\fchirp$. 
Fig.~\ref{f3} shows the (color-coded) analytical result~\eqref{e21}. 
The color-filled circles superimposed to the colored, analytical "background" indicate the corresponding SBE results, $\Delta\omegapeaksbe$; the circle-colors follow the same scale abopted also for the analytical data. 

In Fig.~\ref{f3}, we observe that SBE peak shifts (circles) are in good overall quantitative agreement with the analytical prediction (background): the color of the circular discs matches the background. (Averaging over the entire plane, we compute a mean absolute deviation of only 0.21\,$\omega_0$.)

At weak chirp and small peak frequencies (low harmonics), the SBE-simulations exhibit many vertical maximum-intensity lines; the corresponding peak shifts vanish. 
These vanishing peak shifts appear in a   region in the phase diagram Fig.~\ref{f3}, which reveals itself as the area that supports light gray circles. 
The region has a characteristic boundary corresponding to 
$|\Delta\omegapeak|\letext0.5\,\omega_0$; 
it is indicated by dashed lines in Fig.~\ref{f3}. Within this region, the relative discrepancy to our analytical formula~\eqref{e21} is somewhat enhanced.
Outside this region, we find the mean relative absolute deviation in the peak shift to be only 17\,\% between SBE simulations and the analytical formula~\eqref{e21}.
\section{comparison of the analytical formula to experiments}
\begin{figure}[]
    \centering
 	\includegraphics[width=8.6cm]{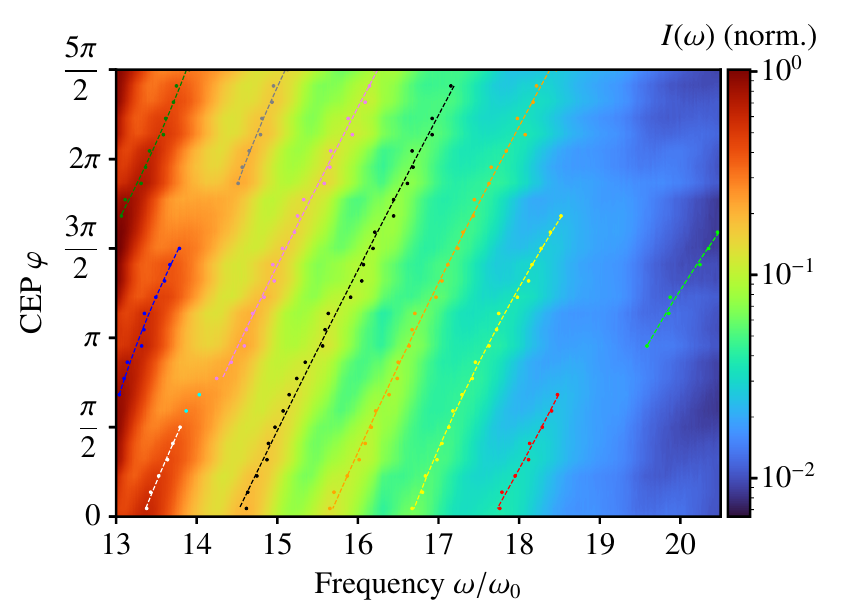}
    \caption{CEP-dependency of the  high-harmonics spectrum of Bi$_2$Te$_3$; experimental data taken from Ref.~\cite{Schmid2021}.
We report local maxima as colored dots. 
We form sets of local maxima, as indicated by the various colors. 
For each set, we perform a quadratic fit~$\omega(\varphi) \eqt\omega_0[ \alpha\pt\beta(\varphi\mt\bar\varphi)\pt\gamma(\varphi\mt\bar\varphi)^2]$ (dotted lines, $\bar\varphi$ is fixed as the average CEP of a line segment, fit parameters~$\alpha,\beta,\gamma$ are reported in Tables~\ref{t1},\ref{t2}).
    \label{f4}}
\end{figure}

We compare our findings~\eqref{e19} and~\eqref{e21} to the experimental high-harmonics spectra emitted from the topological surface state of Bi$_2$Te$_3$~\cite{Schmid2021}, reproduced in Fig.~\ref{f4}. This data displays the characteristic stripe pattern that our theoretical analysis predicts. 
Beyond this, there is also a qualitative agreement in details; e.g., the increase of the tilt angle with growing harmonic order predicted in \eqref{e19} is also seen in Fig.~\ref{f4}.
For a quantitative analysis, we fit parabol{\ae}
\begin{align}
\omega(\varphi) \eqt\omega_0[ \alpha\pt\beta(\varphi\mt\bar\varphi)\pt\gamma(\varphi\mt\bar\varphi)^2]
\end{align}
to discrete local maxima with fit parameters $\alpha,\beta,\gamma$ and fixed $\omega_0,\bar\varphi$, see Fig.~\ref{f4} and caption.
We find that the linear term~$\beta$ (reported in Table~\ref{t1}) is dominating the fit, in line with our analytical result~\eqref{e19}.
From our quantitative analysis, we also find that tilt angles tend to increase with the frequency, see Table~\ref{t1}.
From 14th to 19th order we observe a "locking", i.e., the average tilt angle between 15th and 18th harmonic order changes within 3\,\%, only (Table~\ref{t1}).
This locking is a manifestation of an equidistant placement of the maxima. They reflect combined properties of pulse shape and band structure, that are not captured in our simplified analytical model, but prevail in SBE simulations. Hence, it is not surprising that locking effects also appear in  Fig.~\ref{f2}.

The linear coefficient $\beta$ together with the  analytical result \eqref{e19} provides an estimate for the pulse-shape parameter, $2\pi \mffbarprime/\omega_0\eqt {-}\sdis 0.067 \pm 0.001$.
The experimental pulse shape has been reported in Ref.~\cite{Schmid2021}, so that 
$\mffbarprime$ can be directly calculated for the given pulse as $2\pi \mffbarprime/\omega_0\apt -0.037$, see App.~\ref{a6}; this is half the fitted value. Given that the experimental pulse shape is parametrically not even close to the regime of applicability of the analytical formula \eqref{e19} --  
the experimental curvature $\mff''(t)$ is far from negligible, see App.~\ref{a6} for a detailed analysis -- 
we find the semi-quantitative agreement encouraging. 

So far, our focus has been on topological surface states of 3D topological insulators. 
Ref.~\cite{Schmid2021} reports harmonic orders up to 13th emananating also from the semiconducting bulk of the topological insulator Bi$_2$Te$_3$. 
We now compare our analytical results with this experiment.
For the pulse shape used in the bulk measurement, we evaluate a tiny chirp $2\pi \mffbarprime/\omega_0\apt {-}\,0.007$ (cf. App.~\ref{a6}).
Inserting that chirp into our CEP shift formula~\eqref{e19}, we predict a slope $d\omega/d\varphi\eqt0.11\omega_0/(2\pi)$ for $\omega\eqt8\omega_0$. 
This shift is about 5\% of the shift for the topological surface state, which is qualitatively consistent with the experiment: indeed, a small, but non-vanishing slopes can be identified close to the eighth harmonic order~\cite{Schmid2021}. 

 \begin{table}
 \caption{Fit parameter~$\beta$ (tilt angle) of fits reported in Fig.~\ref{f4}, as function of the average frequency~$\omegabar$ of the line segment.
 We report the full set of fit parameters in  App.~\ref{a5}.
 }
 \begin{ruledtabular}
    \begin{tabular}{lcccccccccc}
        $\omegabar/\omega_0$ &  
                13.4 & 13.5 & 13.6 & 14.8 & 15.3 & 15.8 & 17.0 & 17.5 & 18.1 & 20.1
        \\
        $2\pi\beta$ &
                1.81 & 2.00 & 1.89 & 1.86 & 2.23 & 2.25 & 2.19 & 2.24 & 2.39 & 2.84
    \end{tabular}
 \end{ruledtabular}
     \label{t1}
 \end{table}

 \section{Conclusion}
We propose an analytical theory for the carrier envelope phase (CEP) dependency of high-harmonic generation under illumination of a material with strong laser pulses. The central result is a simple analytical formula describing the shifts of high-harmonic peaks under the change of the CEP. This formula explains, e.g., why peak positions can occur at non-integer harmonic orders. Further, it predicts that the shift velocity is proportional to the peak frequency {and the chirp of the driving laser pulse}. 
The comparison with a full-fledged simulation based on the semiconductor-Bloch formalism establishes the quantitative accuracy of the analytical result in a large parameter regime. Also the comparison to the experiment~\cite{Schmid2021} is surprisingly favorable given that the experimental pulse shape is only marginally consistent with the conditions of applicability of our formula.  

Our theory provides the first understanding of the phenomenon of CEP shifts in materials based on simple, analytically derived parametric dependencies.
We conclude emphasizing the broad applicability of our result, 
the validity of which we have demonstrated for a large parameter regime and a wide range of material classes. Our work represents another stepping stone towards understanding the microscopic mechanisms underlying high-harmonic generation in materials. 

\section*{Code availability}

For all SBE simulations, we have used our program package CUED~\cite{Wilhelm2021}, that is freely available, \url{https://github.com/ccmt-regensburg/CUED}.

\begin{acknowledgments}
We thank P.~Grössing, C.~Schmid, M.~Stefinger, and L.~Weigl for helpful discussions. 
We acknowledge support from the German Research Foundation (DFG) through the Collaborative Research Center, SFB 1277 (project A03) and through the State Major Instrumentation Programme, No.~464531296.
The authors gratefully acknowledge the Gauss Centre for Supercomputing e.V. (www.gauss-centre.eu) for funding this project by providing computing time on the GCS Supercomputer SuperMUC-NG at Leibniz Supercomputing Centre (www.lrz.de) via project pn72pa.
\end{acknowledgments}

\section*{Author contributions}
M.G. carried out SBE simulations and analyzed the numerical data.
M.N, M.G., F.E., and J.W. developed the analytical model.
A.S. developed the SBE simulation code. 
F.E. and J.W. conceived the study, supervised the project and wrote the paper with contributions from all authors. 
All authors contributed to discussing the results.

\appendix

\section{Computational details and convergence tests}\label{app:convergencetests}
\begin{figure}[b]
    \centering
    \includegraphics[width=8.6cm]{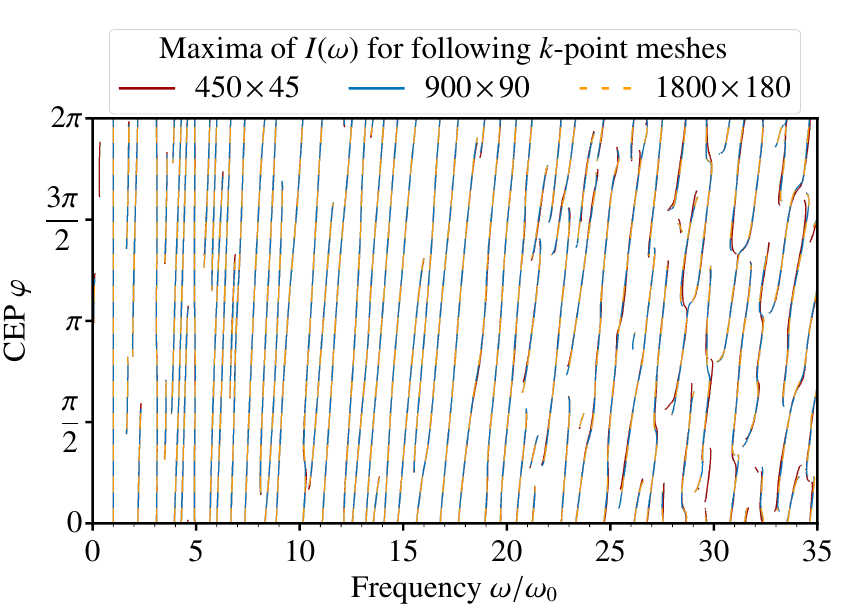}
    \caption{
    Maxima of $I(\omega)$ computed with parameters as in Fig.~\ref{f2}\,c for three different Monkhorst-Pack $k$-point meshes~\cite{Monkhorst1976}  $N_1\timest N_2$ ($450\timest45,900\timest90,1800\timest180$), where $N_1$ is the number of $k$-points in $\Gamma$-M direction and $N_2$ is the number of $k$-points orthogonal to the $\Gamma$-M direction. 
}
    \label{fig - convergence}
\end{figure}

As electric field, we  use Eq.~\eqref{eq-Efield} throughout where the parameters $E_0\eqt3\,\mathrm{MV/cm}$, $\omega_0\eqt2\pi\cdott25\,\mathrm{THz}$ and $\sigma\eqt90\,\mathrm{fs}$ were fixed for the calculations in the main text.
We align the electric field along the $\Gamma$-M direction which we label as $x$-direction in Eq.~\eqref{eq-Efield}. 
We have used the two-band Dirac-like model Hamiltonian for the topological surface states of  Bi$_2$Te$_3$ from Ref.~\cite{Schmid2021} together with a hexagonal Brillouin zone with a size that stems from a real space lattice constant $a\eqt4.396\,\AA$.
As in Ref.~\cite{Schmid2021}, we start from an equilibrium band occupation that is given by a Fermi-Dirac distribution with a Fermi level of  0.176\,eV above the conduction band minimum and with a temperature of 30\,meV.
We compute the time-dependent density matrix $\rho(t)$ from SBE in the velocity gauge with a dephasing time $T_2\eqt10$\,fs which is an accepted simulation value~\cite{Floss2018}. 
For the time evolution in the SBE formalism, we employ an adaptive algorithm~\cite{Virtanen2020} with a maximum time step of $0.1$\,fs and with a time window of  $[-500\,\text{fs},500\,\text{fs}]$.
These settings lead to intensity spectra that are converged with respect to time discretization.
The Fourier transform to frequency domain includes a Gaussian window function with full width at half maximum of $2\sqrt{\ln 2}\cdot90$\,fs.
In the SBE, dipoles are used which are diverging for the Dirac-like two-band Hamiltonian at the $\Gamma$-point~\cite{AlNaib2014,Wilhelm2021}. 
Thus, we carefully checked the convergence of the $k$-point mesh, see  Fig.~\ref{fig - convergence}. 
We observe excellent agreement between the $900\timest90$ and $1800\timest180$ $k$-mesh.
We conclude that the $900\timest90$ mesh is sufficient to reach convergence in the $k$-point mesh size and therefore, we have used a $900\timest90$ mesh  for all SBE calculations. 
In all figures, we have varied the CEP from 0 to $2\pi$, where we have used $N_\text{CEP}\eqt 384$ discrete CEPs in the $[0,2\pi]$ window thoughout. 

\section{Fourier transform of the time-dependent current \label{Appendix-Fourier-trafo}}
The time derivative of the current~\eqref{eq-current} is
\begin{align}
   \partial_t\bv(t)  = -\,2\,\hat{\mathbf{x}}\,\frac{e\vF}{\hbar} \, E(t) \;\delta(\sdis k_x(t)\sdis )\,,\label{eq-jt}
\end{align}
where $\delta$ denotes the Dirac delta function. 
Then,
\begin{align}
    \delta(\sdis k_x(t)\sdis ) &= \sum_m \frac{\delta(t-t_m(\varphi))}{\left|\left. \partial_t k_x(t) \right|_{t_m(\varphi)} \right|}
    =
    \sum_m \frac{\delta(t-t_m(\varphi))}{e \left|E(t_m(\varphi))\right|/\hbar}\,.\label{eq-deltasum}
\end{align}
Combining Eqs.~\eqref{eq-jt} and~\eqref{eq-deltasum}, we arrive at Eq.~\eqref{eq-current-ft},
\begin{align}
     \bv(\omega) = \frac{1}{i\omega}\,\partial_t\bv(\omega)=\frac{2\vF}{i\omega } \;\hat{\mathbf{x}}
     \sum_m (-1)^{m+1} \exp[i\omega t_m(\varphi)]
    \,.
\end{align}
The linear dispersion~\eqref{eq-lindisp} of the model band structure is justified by the Dirac character of the surface conduction band of Bi$_2$Te$_3$ close to the $\Gamma$-point~\cite{Liu2010}.
For the other commonly adopted model band structure, a parabolic dispersion $\epsilon(\bk)\eqt \hbar^2|\bk|^2/(2m)$, no high-harmonic emission is observed under driving by an electric field from Eq.~\eqref{eq-Efield}.
This is due to the velocity~$\bv(t)\eqt \hbar\bk(t)/m$ [Eq.~\eqref{eq-current-quasiclassical}] oscillating solely with the fundamental frequency~$\omega_0$. 
We also do not consider excitonic effects and other electron-electron interaction during the non-equilibrium dynamics as they are believed to have negligible contributions~\cite{Schmid2021}.
Also, we omit bulk bands which have been shown to not contribute to the high-harmonic emission for the pulse shape we consider in this work~\cite{Schmid2021}.
\section{General discussion of CEP-shifts at weak chirp \label{a4}}

We give more details on the discussion of the spacing of roots in Sec.~\ref{secIV}. 
The starting point is a homogeneous spacing, $\tO_m(\varphi)\eqt (m\pi\mt\varphi)/\omega_0$ of the zeros of~\eqref{e8}, $\bk(\tO_m(\varphi))\eqt0$, where $\omega_0$ is the fundamental frequency.
We achieve a non-uniform spacing by implementing a small rescaling of the time, $\bk(t)\rightarrowtext\bk\big(t(1\pt \mff(t)\big)$, such that we have for the zeros~$t_m(\varphi)$ with non-uniform spacing
\begin{align}
  t_m^{(0)}(\varphi) = \Big[\sdis 1+\mff(t_m(\varphi))\sdis\Big]\,t_m(\varphi) \,.
   \label{eC1} 
\end{align}
The chirp introduced in Sec.~\ref{secIV} corresponds to a linear dependency $\mff(t)\eqt \fchirp t$. We consider more general situations subject to the condition that $\mff(t)$ is slowly varying from one zero to the next. Solving \eqref{eC1} for $t_m$ we have
\begin{align}
    t_m(\varphi) &= \tO_m(\varphi) -\mff(\tO_m(\varphi))\ \tO_m(\varphi) + \mathcal{O}(\mff^2) \nonumber\\
    &\approx\tO_m(\varphi)\left[1-\mff(\tO_m(\varphi))\right] \label{ec2}
\end{align}
where gradient terms have been neglected. Further, after defining the equal spacing
($\tO_{\ell m}\coloneqqt \tO_\ell(\varphi) \mt \tO_m(\varphi)\eqt\pi(\ell\mt m)/\omega_0\eqt \tO_{\ell-m}(\varphi{=}0)$), we  define the non-uniform spacing~$t_{\ell m}(\varphi)\coloneqqt t_\ell(\varphi) \mt t_m(\varphi)$ 
\begin{align}
   t_{\ell m}(\varphi) &\approx  \tO_{\ell m} - 
    \left[ 
    \tO_\ell(\varphi)\,\mff(\tO_\ell(\varphi))- 
    \tO_m(\varphi)\,\mff(\tO_m(\varphi))
    \right] \nonumber \\
    &\approx \tO_{\ell m}\left[ 1 - \mff(\TO_{\ell m}(\varphi)) - \TO_{\ell m}(\varphi)\, \mff'(\TO_{\ell m}(\varphi)) \right] 
\end{align}
where $\TO_{\ell m}(\varphi)\coloneqqt(\tO_\ell(\varphi)\pt \tO_m(\varphi))/2 \eqt \tO_{(\ell+m)/2}(\varphi)$  and terms involving second derivatives have been dropped. Similarly, we have  
\begin{align}
    \partial_\varphi t_{\ell m} 
    &= 2 \tO_{\ell m}\, \mff'(\TO_{\ell m})  /\omega_0 \label{eq-app-dtdphi}
\end{align}
 and Eqs.~\eqref{e17} and~\eqref{e18} are therefore to leading order in $\mff$ 
\begin{align}
    I_t(\omega)& =  \sum_{\ell,m}^{N_z}(-1)^{\ell+m} e^{i\omega \tO_{\ell m}}\ \tO_{\ell m}\,,
    \\
    I_t^\prime(\omega,\varphi) &= \frac{2}{\omega_0} \sum_{\ell,m}^{N_z}(-1)^{\ell+m} e^{i\omega \tO_{\ell m}}\ \tO_{\ell m} \  \mff'(\TO_{\ell m})  \,.
\end{align}
For the situation of an equidistant spacing, the double sum can be reorganized as a sum over pairs. The first summation is over pairs with the same distance, $s\eqt \ell\mt m$, while the second sum is over the different values $M\eqt(m\pt\ell)/2$ that these pairs will have 
\begin{align}
    I_t(\omega)& =  \sum_{s=-N_z+1}^{N_z-1} e^{i\omega \tOO_{s}}\ \tOO_{s}
    \sum_{M=1+|s|/2}^{N_z-|s|/2} (-1)^{2M}\\
    I_t^\prime(\omega,\varphi) &= \sdis\frac{2}{\omega_0} \sum_{s=-N_z+1}^{N_z-1}e^{i\omega \tOO_{s}}\ \tOO_{s}  \sum_{M=1+|s|/2}^{N_z-|s|/2}(-1)^{2M} \ \mff'(\tO_{M}(\varphi))
\end{align}
 where we abbreviated $\tOO_s\coloneqqt \tO_s(\varphi{=}0)\eqt \pi s /\omega_0$. 
Using the relation 
\begin{align}
    \sum_{M=1+|s|/2}^{N_z-|s|/2} (-1)^{2M} = (-1)^{s}(N_z-|s|)
\end{align}
 we can simplify 
\begin{align}
    I_t(\omega)& = \frac{\pi}{\omega_0} \sum_{s=-N_z+1}^{N_z-1} e^{i\pi(\omega /\omega_0+1)s}\, s\, 
    (N_z-|s|)\label{ec11}
    \\
    I_t^\prime(\omega,\varphi) &= \frac{2\pi}{\omega_0^2} \sum_{s=-N_z+1}^{N_z-1}e^{i\pi(\omega /\omega_0+1)s}\,  s\, (N_z-|s|) \ \bar\mff'(s;\varphi)\,; \label{ec12}
\end{align}
in the last line we have introduced an average chirp 
\begin{align}
    \bar \mff'(s;\varphi) %&\coloneqq \frac{1}{(N_z-|s|)}\sum_{M=1+|s|/2}^{N_z-|s|/2}(-1)^{2M-|s|} \ \mff'(\TO_{2M}(\varphi)) \nonumber\\
& \coloneqq \frac{1}{N_z-|s|}\sum_{M=1+|s|/2}^{N_z-|s|/2} \mff'(\tO_{M}(\varphi)) \label{ec14}
\end{align}
motivated by the observation that a factor $(-1)^{2M-|s|}$  in the $M$-summation can be replaced by unity.
In the special situation where $\bar\mff'(s,\varphi)$ is independent of $s$ and $\varphi$, we have $I_t^\prime/I_t \eqt \sdis 2\sdis \bar\mff'/\omega_0$; with Eq.~\eqref{e16}, we obtain our main result~\eqref{e19} \begin{align}
\frac{d\omega}{d\varphi} = -\sdis2\sdis\frac{\omega}{\omega_0}\bar\mff'\,.
\end{align}

In particular, when choosing a sinusoidal pulse~\eqref{eq-Efield}, the zeros of $\bk(t)$ for $\bk_0\eqt 0$ are obtained as
\begin{align}
t_m = \frac{1}{\omega_0} \,((m\sdis{+}\sdis\frac{1}{2})\,\pi-\varphi)\,(1-\frac{\fchirp}{\omega_0}\,((m\sdis{+}\frac{1}{2}\sdis)\,\pi-\varphi)) \,,\label{eq-tm}
\end{align}
in the limit of small  chirp ($|\fchirp| \hspace{0.1em}{\ll}\hspace{0.1em} \omega_0$) and for small $|m|, m\intext \mathbb{Z}$ as it is applicable in case of a few-cycle pulse $E(t)$. When comparing to Eq.~\eqref{ec2}, we obtain $\tO_m(\varphi)\eqt ((m\pt1/2)\pi\mt\varphi)/\omega_0$, $\mff(\tO_m(\varphi))\eqt \fchirp \tO_m(\varphi)$ and $\mffbarprime\eqt\fchirp$.

\section{High-harmonic frequencies for the 1D Dirac dispersion}\label{a4a}
In this section, we analytically calculate the chirp-free high-harmonic peak frequencies of our semiclassical model from Sec.~\ref{secIV}.
Following App.~\ref{a4}, we have at zero chirp ($\mff(t)\eqt0$) an equidistant spacing of roots of the electron excursion~$\bk(t)$ in the Brillouin zone, $t_{\ell m} \eqt\pi(\ell\mt m)/\omega_0$.
Then, the emission intensity~$I(\omega)$ from Eq.~\eqref{e13} turns into
\begin{align}
I(\omega)
     \propto \sum_{\ell,m}^{N_z} (-1)^{\ell+m}e^{i(\ell-m)\omega/\omega_0} \,.
\end{align}
Using the substitution $s\eqt \ell\mt m$ and $M\eqt(m\pt\ell)/2$, we obtain similarly to App.~\ref{a4} the emission intensity
\begin{align}
    I(\omega) \propto
    \Big(2 N_z
    + \frac{\omega_0}{\pi}\frac{\partial}{\partial\omega} \Big)
     \sum_{s=0}^{N_z-1} \cos(\pi s (\omega/\omega_0+1))
\end{align}
which is independent of the CEP~$\varphi$.
In the limit $N_z\rightarrowtext\infty$, the sum takes a non-zero value only for $\omega\eqt (2k\pt1)\,\omega_0, k\intext \mathbb{Z}$.
We recover that high-harmonic peaks appear at odd orders in inversion-symmetric materials.
"Non-integer" HHG is therefore only present for non-vanishing chirp, $\mff(t)\neqt0$, where peak frequencies $\omega$ get shifted according to Eq.~\eqref{e21}.

\section{CEP shifts for generic band structures}\label{app-cepshiftsgenericbands}
We derive Eqs.~\eqref{e19} and \eqref{e21} for a more generic situation, where the current needs to fulfill
\begin{align}
     \partial_t \bv(\omega) =
     \sum_{m=1}^{N_z}  \mathbf{v}_m \exp(i\omega t_m(\varphi))
\end{align}
for high-harmonic frequencies $\omega \ggtext \omega_0$.
The time points~$t_m$ are non-equidistant and are shifted by a linearized slope of $\mff(t) \eqt \fchirp t$ according to
\begin{align}
    t_m(\varphi) =\tO_m(\varphi)\left[1-\fchirp \tO_m(\varphi) \right]
\end{align}
with equidistant $\tO_m(\varphi)\eqt (m\pi\mt\varphi)/\omega_0$.
Starting from our definition of CEP shifts \eqref{e4}, we calculate
\begin{align}
    (\partial I/\partial \omega)_\varphi &\eqt i \sum_{\ell,m}^{N_z} \alpha_{\ell m} e^{i\omega t_{\ell m}} t_{\ell m}\\
    (\partial I/\partial \varphi)_\omega &\eqt i \omega \sum_{\ell,m}^{N_z} \alpha_{\ell m} e^{i\omega t_{\ell m}} \partial_\varphi t_{\ell m}\,,
\end{align}
where we have defined $\alpha_{\ell m} \coloneqqt \mathbf{v}_\ell \cdot \mathbf{v}_m^*$. The non-uniform time spacings $t_{\ell m} \eqt t_{\ell m}(\varphi) \eqt t_\ell(\varphi) - t_m(\varphi)$ can be computed as in App.~\ref{a4}. 
In the spirit of App.~\ref{a4}, we reorganize the sums from indices $\ell,\,m$ to $s\eqt\ell\mt m$ and $M\eqt(m\pt\ell)/2$. By identifying $\alpha_{s M} \eqt \alpha_{\ell m}$ and using Eq.~\eqref{eq-app-dtdphi}, we obtain in leading order in $\fchirp$:
\begin{align}
    (\partial I/\partial \omega)_\varphi &\eqt i \sum_{s=1-N_z}^{N_z-1} \tO_s e^{i\omega\tO_s} \sum_{M=1+|s|/2}^{N_z-|s|/2} \alpha_{s M} \label{eq-app-e5}\\
    (\partial I/\partial \varphi)_\omega &\eqt i\omega \sum_{s=1-N_z}^{N_z-1} \tO_s e^{i\omega\tO_s} \sum_{M=1+|s|/2}^{N_z-|s|/2} \alpha_{s M} \frac{2}{\omega_0} \mff'(\tO_{M}(\varphi)) \label{eq-app-e6}\,,
\end{align}
where we use the equidistant spacing $\tO_{\ell m} \eqt \tO_\ell(\varphi) \mt \tO_m(\varphi)\eqt\pi(\ell\mt m)/\omega_0$ from App.~\ref{a4}.
Considering an electric field \eqref{eq-Efield}, we can simplify the expression $\mff'(\tO_{M}(\varphi)) = \fchirp$ for all $M$. 
We insert Eq.~\eqref{eq-app-e5} and \eqref{eq-app-e6} into Eq.~\eqref{e4}, arriving at our main result \eqref{e19} in a slightly modified form:
\begin{align} 
    \frac{d\omega}{d\varphi} 
    \eqt  - \frac{(\partial I/\partial \varphi)_\omega}{(\partial I/\partial \omega)_\varphi}
    \eqt - 2\frac{\omega}{\omega_0}\fchirp
    \,. 
\end{align}

\begin{table*}
 \caption{Fit parameters~$\alpha,\beta,\gamma$ of the fits $\omega(\varphi)\eqt\omega_0[ \alpha\pt\beta(\varphi\mt\bar\varphi)\pt\gamma(\varphi\mt\bar\varphi)^2]$ reported in Fig.~\ref{f4}, as function of the average frequency~$\omegabar$ of the line segment.
 We also report the average CEP~$\bar\varphi$ of the line segments. 
 }
 \begin{ruledtabular}
    \begin{tabular}{lcccccccccc}
        $\omegabar/\omega_0$ &  
                13.4 & 13.5 & 13.6 & 14.8 & 15.3 & 15.8 & 17.0 & 17.5 & 18.1 & 20.1
                \\[0.2em]
        $\alpha$ &
                --\,0.02 & 0.01 & 0.00 & --\,0.01 & --\,0.01 & --\,0.01 & --\,0.03 & --\,0.04 & 0.00 & --\,0.01
                \\[0.2em]
        $2\pi\beta$ &
                1.81 & 2.00 & 1.89 & 1.86 & 2.23 & 2.25 & 2.19 & 2.24 & 2.39 & 2.84
                \\[0.2em]
        $\gamma$ &
                0.02 & --\,0.02 & 0.01 & 0.02 & 0.00 & 0.00 & 0.00 & 0.02 & --\,0.01 & 0.03
                \\[0.2em]
        $\bar\varphi/\pi$ &
                1.1 & 2.1 & 0.3 & 2.2 & 1.6 & 1.2 & 1.3 & 0.9 & 0.4 & 1.3
    \end{tabular}
 \end{ruledtabular}
     \label{t2}
\end{table*}

\begin{figure*}
    \centering
    \includegraphics[width=17.6cm]{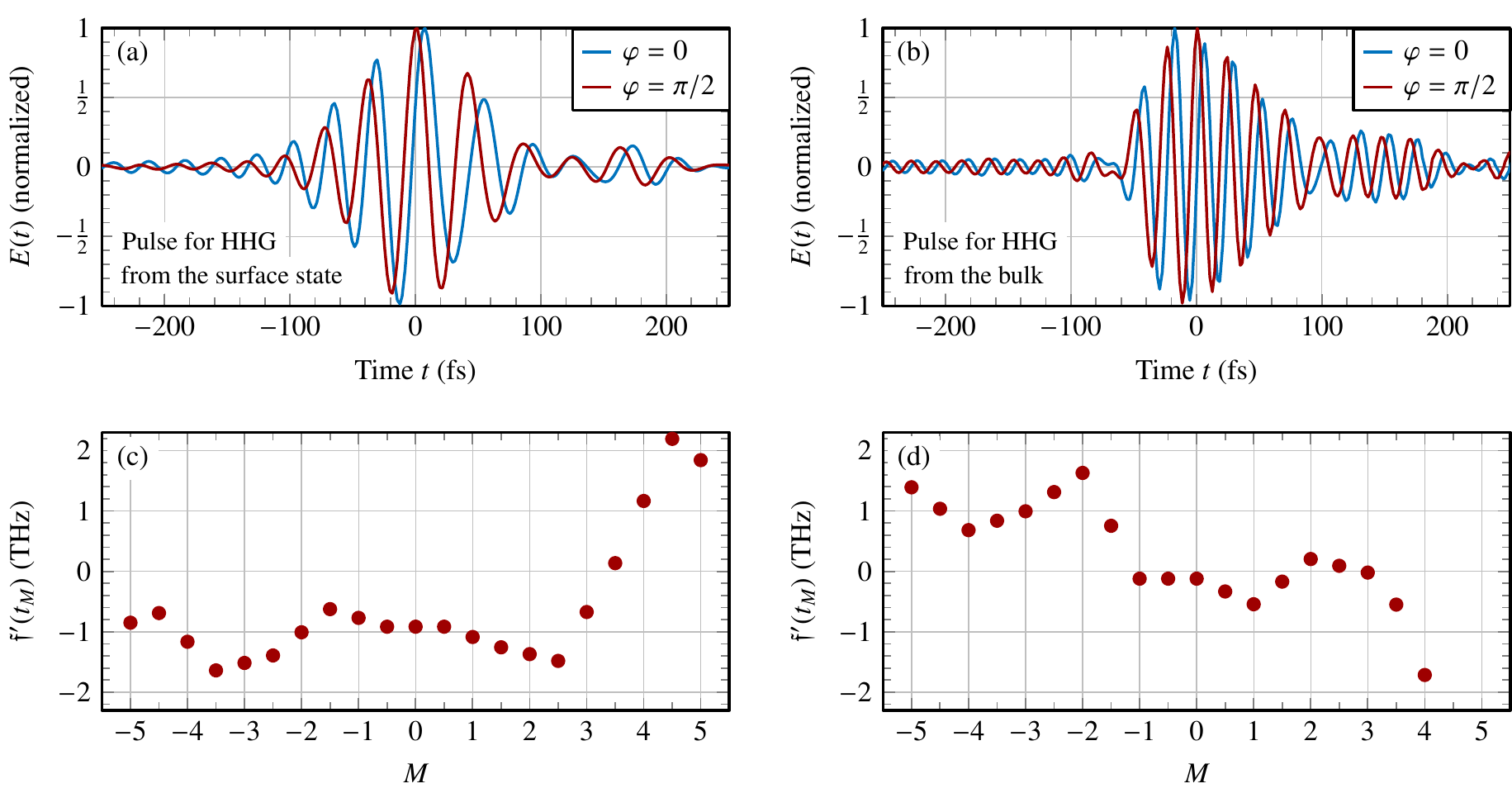}
    \caption{
     (a) and (b): Experimental pulse shapes from Ref.~\cite{Schmid2021} that lead to HHG (a) from the topological surface state of Bi$_2$Te$_3$ and (b) from the bulk of Bi$_2$Te$_3$. (c) and (d): Time-local chirp~$\mff'(\tO_M(\varphi{=}\pi/2))$ evaluated from Eq.~\eqref{ec2} for the pulse shapes from (a) and (b), respectively.
    At highest field, $M\eqt0$, we have (c) $\mff'(\tO_0(\varphi{=}\pi/2))\eqt {-}\,1.0$\,THz and (d) $\mff'(\tO_0(\varphi{=}\pi/2))\eqt {-}\,0.1$\,THz.}
    \label{f6}
\end{figure*}
\section{Parameters of the fits reported in Fig.~\ref{f4}}\label{a5}
In Table~\ref{t2}, we report the full set of fit parameters obtained from the fitting in Fig.~\ref{f4}.

\section{Evaluating the time-local chirp of experimental electric field pulses}
\label{a6}
In this appendix, we evaluate the average chirp~$\mffbarprime$ [definition in Eq.~\eqref{ec14}] for two electric field pulses that have been employed in the experiment by Schmid \textit{et al.}~\cite{Schmid2021}
The experimental electric field pulses are sketched in Fig.~\ref{f6}\,(a) and~(b), where pulse (a) has been used for HHG from the topological surface state of Bi$_2$Te$_3$ and pulse (b) for HHG from the bulk of Bi$_2$Te$_3$~\cite{Schmid2021}. 
We evaluate the "time-local chirp" $\mff'(\tO_{M}(\varphi)) $ which is the key quantity in our analytical formulae~\eqref{e19} and~\eqref{e21}.
$\mff'(\tO_{M}(\varphi)) $ follows from Eq.~\eqref{ec2}, $\mff'(\tO_{M}(\varphi)) \eqt 1\mt t_m(\varphi)/\tO_m(\varphi)$.
The results for $\mff'(\tO_{M}(\varphi)) $ for the pulses in (a) and (b) with CEP $\varphi\eqt\pi/2$ are sketched in Fig.~\ref{f6}\,(c) and~(d), respectively.
For the pulse from Fig.~\ref{f6}\,(a), we observe that close to $M\eqt0$, we have a constant time-local chirp $\mff'\apt{-}\sdis 0.92$\,THz (giving $-2\pi \mffbarprime/\omega_0\apt 0.037$ with $\omega_0\eqt2\pi\cdott25$\,THz) while for roots $|M|\gt1$, the time-local chirp varies. 
Thus, higher-order derivatives of $\mff$ become important that are not included in the analysis in App.~\ref{a4} and in our analytical result~\eqref{e19}. 
For the pulse from Fig.~\ref{f6}\,(b), we calculate an average time-local chirp $\mffbarprime\eqt 0.29$\,THz (average taken in the time interval $[-40\,\text{fs}, 50\,\text{fs}]$). 
With the frequency $\omega_0\eqt2\pi\cdott42$\,THz, we have $-2\pi\mffbarprime/\omega_0\eqt 0.0069$ for the pulse from Fig.~\ref{f6}\,(b).

Please note, that, compared to the data in Ref.~\citenum{Schmid2021}, we have redefined the CEP, $-\,\varphi \rightarrowtext \varphi\pt2\pi$, to match the definition in Ref.~\cite{Schmid2021} to the definition in our work. 

\section{Verification of the analytical CEP-shift formula for various pulse shapes and Hamiltonians}\label{a9}
\begin{figure*}
        \centering
    	\includegraphics[width=16cm]{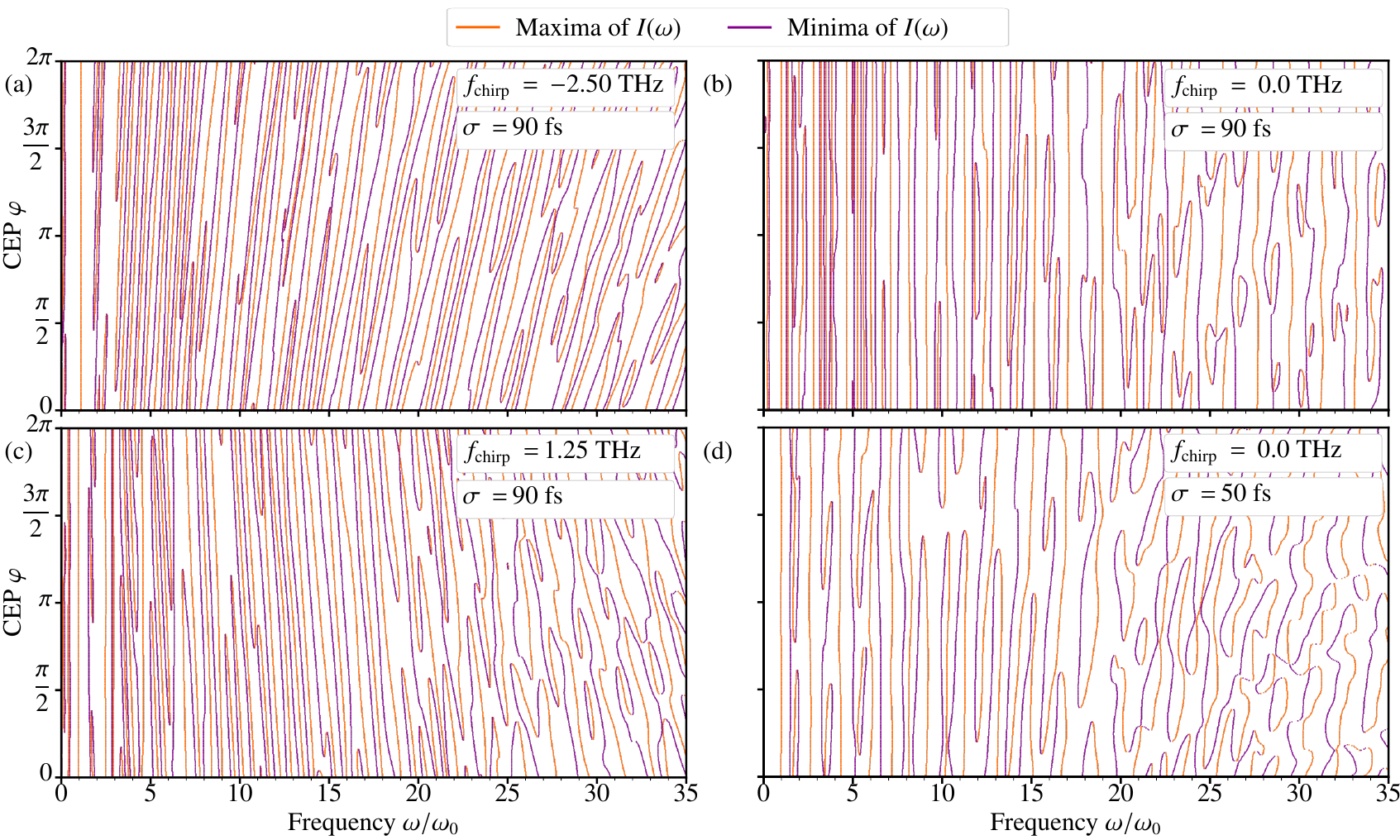}
        \caption{
        Local extrema of~$I(\omega)$ computed from SBEs for the  Bi$_2$Te$_3$-surface-state Hamiltonian as in Fig.~\ref{f2}\,(b) but with varying pulse duration and with varying chirp in the driving electric field~\eqref{eq-Efield}:
(a) $\fchirp\eqt{-}2.5$\,THz, $\sigma\eqt$90\,fs, 
(b) $\fchirp\eqt0$, $\sigma\eqt$90\,fs, 
(c) $\fchirp\eqt1.25$\,THz, $\sigma\eqt$90\,fs, 
(d) $\fchirp\eqt0$, $\sigma\eqt$50\,fs, 
}
        \label{fig-skeletons}
\end{figure*}

\begin{figure*}    \centering
    \includegraphics[width=16cm]{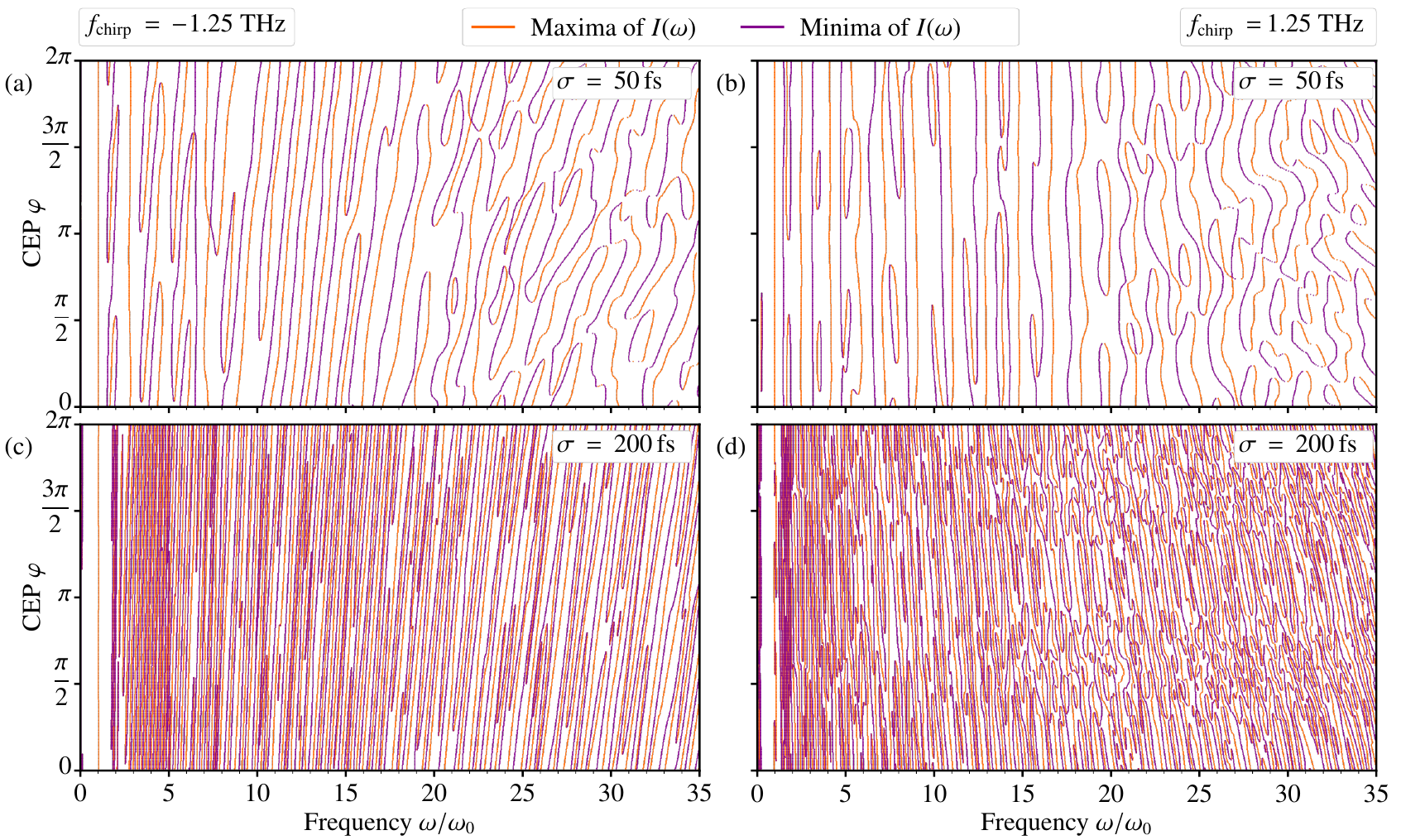}
    \caption{        
    Local extrema of~$I(\omega)$ computed from SBEs for the  Bi$_2$Te$_3$-surface-state Hamiltonian as in Fig.~\ref{f2}\,(b) but with varying pulse duration (a), (b) $\sigma\eqt 50\,\mathrm{fs}$ and (c), (d) $\sigma\eqt 200\,\mathrm{fs}$ (in Fig.~\ref{f2}\,(b): $\sigma\eqt 90$\,fs).
    For (a) and (c), we choose the shape of~$\bE(t)$ [Eq.~\eqref{eq-Efield}] as in Fig.~\ref{f2}\,(b), in particular $\fchirp\eqt-1.25$\,THz. 
    For (b) and (d),  we choose $\fchirp\eqt1.25$\,THz which leads to an inversion of the tilt angle, fully in line with our analytical formula Eq.~\eqref{e19}.\vspace{-2em}
    }
    \label{fig-Sigmacomp}
\end{figure*}

In the main text, we have reported high-harmonics spectra for a time-dependent electric field $E(t)$, Eq.~\eqref{eq-Efield},  with amplitude $E_0\eqt3\,\mathrm{MV/cm}$, frequency~$\omega\eqt 2\pi\cdot 25\,$THz, and pulse duration $\sigma\eqt90\,\mathrm{fs}$.
We also kept the dephasing time $T_2\eqt10\,\mathrm{fs}$ and the Bi$_2$Te$_3$ surface-state Hamiltonian~\cite{Schmid2021} unchanged in all simulations in the main text. 
In the main text, we only varied the CEP~$\varphi$ and the chirp~$\fchirp$. 
In this appendix, we show high-harmonics spectra for more values of $\fchirp$, $\sigma$, $E_0$, $\omega$, $T_2$ and for additional Hamiltonians. 
First, we show the extrema of the emission spectrum for chirp $\fchirp\eqt{-}\,2.5$\,THz, $\fchirp\eqt 0$, and
$\fchirp\eqt 1.25$\,THz   in Fig.~\ref{fig-skeletons}\,(a)\,--\,(c).
We observe that the tilt angle~$d\varphi/d\omega$ for chirp $\fchirp\eqt{-}\,2.5$\,THz [Fig.~\ref{fig-skeletons}\,(a)] is roughly doubled compared to $\fchirp\eqt{-}\,1.25$\,THz (Fig.~\ref{f2}), fully in line with our analytical formula~\eqref{e19}.
For a chirp $\fchirp\eqt{+}\,1.25$\,THz [Fig.~\ref{fig-skeletons}\,(c)], the tilt angle~$d\varphi/d\omega$ changes its sign, i.e.~the extremal lines are tilted to the left (from north west to sourth east), as predicted by our analytical formula~\eqref{e19}.
For chirp $\fchirp\eqt0$, our analytical formula~\eqref{e19} predicts a vanishing tilt angle~$d\varphi/d\omega$. 
This prediction is in line with the simulation reported in Fig.~\ref{fig-skeletons}\,(b) for $\fchirp\eqt0$, where we observe a tilt angle~$d\varphi/d\omega$ that is much reduced compared to $\fchirp\eqt{-}\,2.5$\,THz and 
$\fchirp\eqt 1.25$\,THz [Fig.~\ref{fig-skeletons}\,(a) and~(c)].
We next consider a short pulse with a duration~$\sigma\eqt50$\,fs  that has approximately only a single cycle $\sigma\omega/(2\pi)\eqt 1.25$ (with $\omega\eqt2\pi\cdott25\,$THz). 
Moreover, we choose a vanishing chirp~$\fchirp\eqt0$, thus our analytical formula predicts a vanishing tilt angle.
We report the extrema of the emission spectrum in Fig.~\ref{fig-skeletons}\,(d). 
For harmonics up to 20th order, we observe a small tilt angle $d\omega/d\varphi$, that increases with~$\omega$. 
We observe that irregular patterns arise above 20th harmonic order with positive and negative tilts~$d\omega/d\varphi$.
This pattern hints to another mechanism underlying the CEP shifts which is not related to the chirp of the pulse, that is zero.
Such mechanisms have already been suggested~\cite{You2017}.
We continue to discuss the extrema in HHG for a short pulse~$\sigma\eqt50\,\text{fs}$ with chirp $\fchirp\eqt{\pm}1.25$\,THz. 
The corresponding extrema in the HHG spectra are shown in Fig.~\ref{fig-Sigmacomp}\,(a) and~(b). 
We observe that, the tilt $d\omega/d\varphi$ is to the right for $\fchirp\eqt{-}1.25$\,THz in Fig.~\ref{fig-Sigmacomp}\,(a), in line with the analytical formula~\eqref{e19}.
In contract, no preferred tilt direction is observed for $\fchirp\eqt{+}1.25$\,THz in Fig.~\ref{fig-Sigmacomp}\,(b), which is in contrast to our analytical formula~\eqref{e19}.
We speculate that an additional mechanism for generating CEP shifts are present for such a short pulse with only $\sigma\omega/(2\pi)\eqt 1.25$ cycles, limiting the predictive accuracy of our analytical formula~\eqref{e19} to many-cycle pulses with $\sigma\omega/(2\pi)\ggtext 1$.
For very long pulses with $\sigma\eqt200$\,fs, we  find tilts~$d\omega/d\varphi$ to the right for $\fchirp\eqt{-}1.25$\,THz [Fig.~\ref{fig-Sigmacomp}\,(c)] and tilts to the left for $\fchirp\eqt{+}1.25$\,THz [Fig.~\ref{fig-Sigmacomp}\,(d)]

\begin{figure*}
    \centering
    \includegraphics[width=16cm]{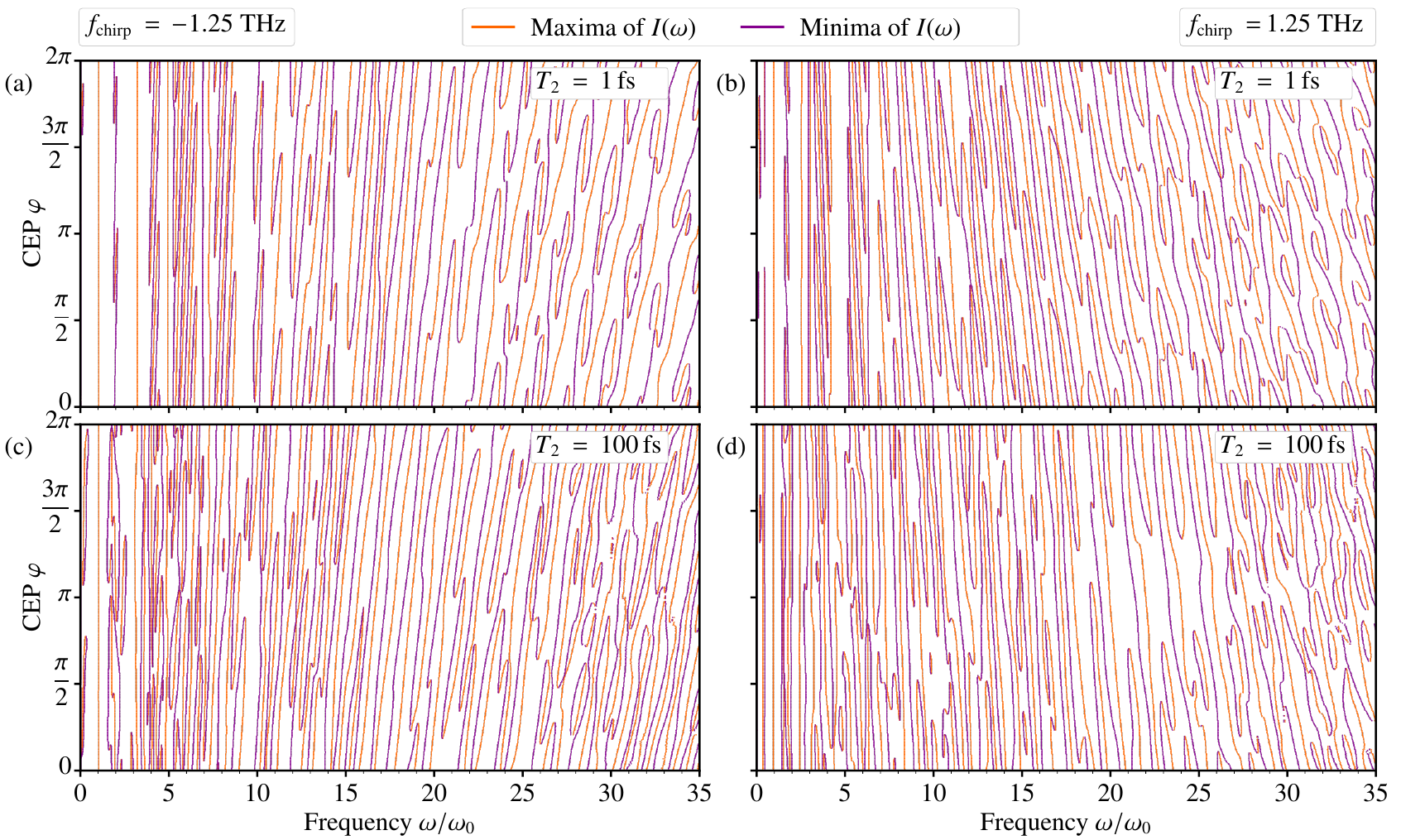}
    \caption{
    Local extrema of~$I(\omega)$ computed from SBEs for the  Bi$_2$Te$_3$-surface-state Hamiltonian as in Fig.~\ref{f2}\,(b), but with a varying dephasing time of (a), (b) $T_2\eqt 1$\,fs and (c), (d) $T_2\eqt 100$\,fs (dephasing time in Fig.~\ref{f2}\,(b): $T_2\eqt10$\,fs).
    For (a) and (c), we choose the driving field~$\bE(t)$ [Eq.~\eqref{eq-Efield}] as in Fig.~\ref{f2}\,(b), in particular $\fchirp\eqt-1.25$\,THz. 
    For (b) and (d),  we choose $\fchirp\eqt1.25$\,THz which leads to an inversion of the tilt angle, fully in line with our analytical formula Eq.~\eqref{e19}.
    }
    \label{fig-T2comp}
\end{figure*}

\begin{figure*}
    \centering
    \includegraphics[width=16cm]{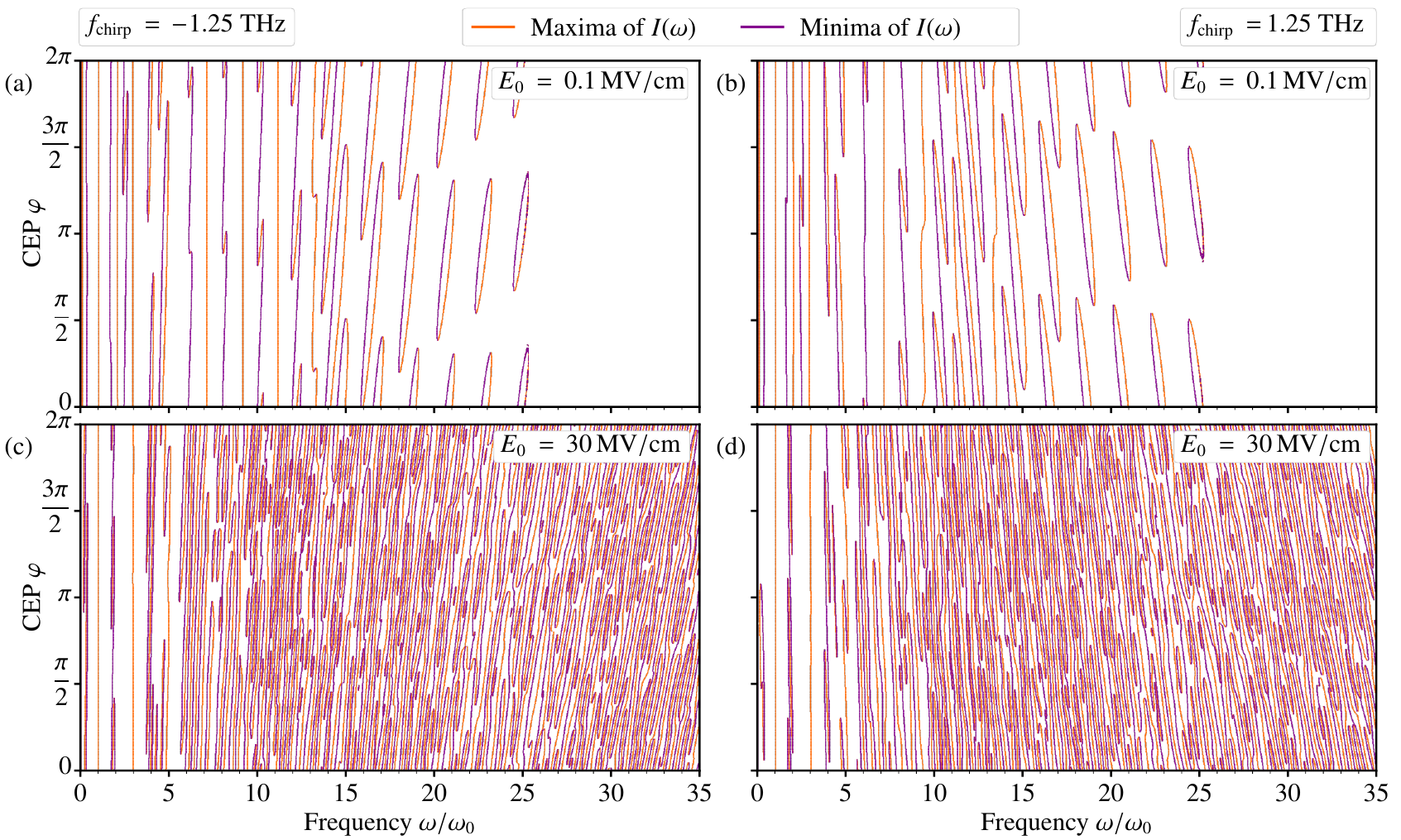}
    \caption{
        Local extrema of~$I(\omega)$ computed from SBEs for the  Bi$_2$Te$_3$-surface-state Hamiltonian as in Fig.~\ref{f2}\,(b) but with varying driving field amplitude (a), (b) $E_0\eqt 0.1\,\mathrm{MV/cm}$ and (c), (d) $E_0\eqt 30\,\mathrm{MV/cm}$ (in Fig.~\ref{f2}\,(b): $E_0\eqt 3$\,MV/cm).
    For (a) and (c), we choose the shape of~$\bE(t)$ [Eq.~\eqref{eq-Efield}] as in Fig.~\ref{f2}\,(b), in particular $\fchirp\eqt-1.25$\,THz. 
    For (b) and (d),  we choose $\fchirp\eqt1.25$\,THz which leads to an inversion of the tilt angle, fully in line with our analytical formula Eq.~\eqref{e19}.
    We show extremal lines in (a) and (b) only up to harmonic order $\omega/\omega_0\eqt26$ due to numerical noise present in higher harmonic orders.\vspace{-2em}
    }
    \label{fig-E0comp}
\end{figure*}

\begin{figure*}
    \centering
    \includegraphics[width=18cm]{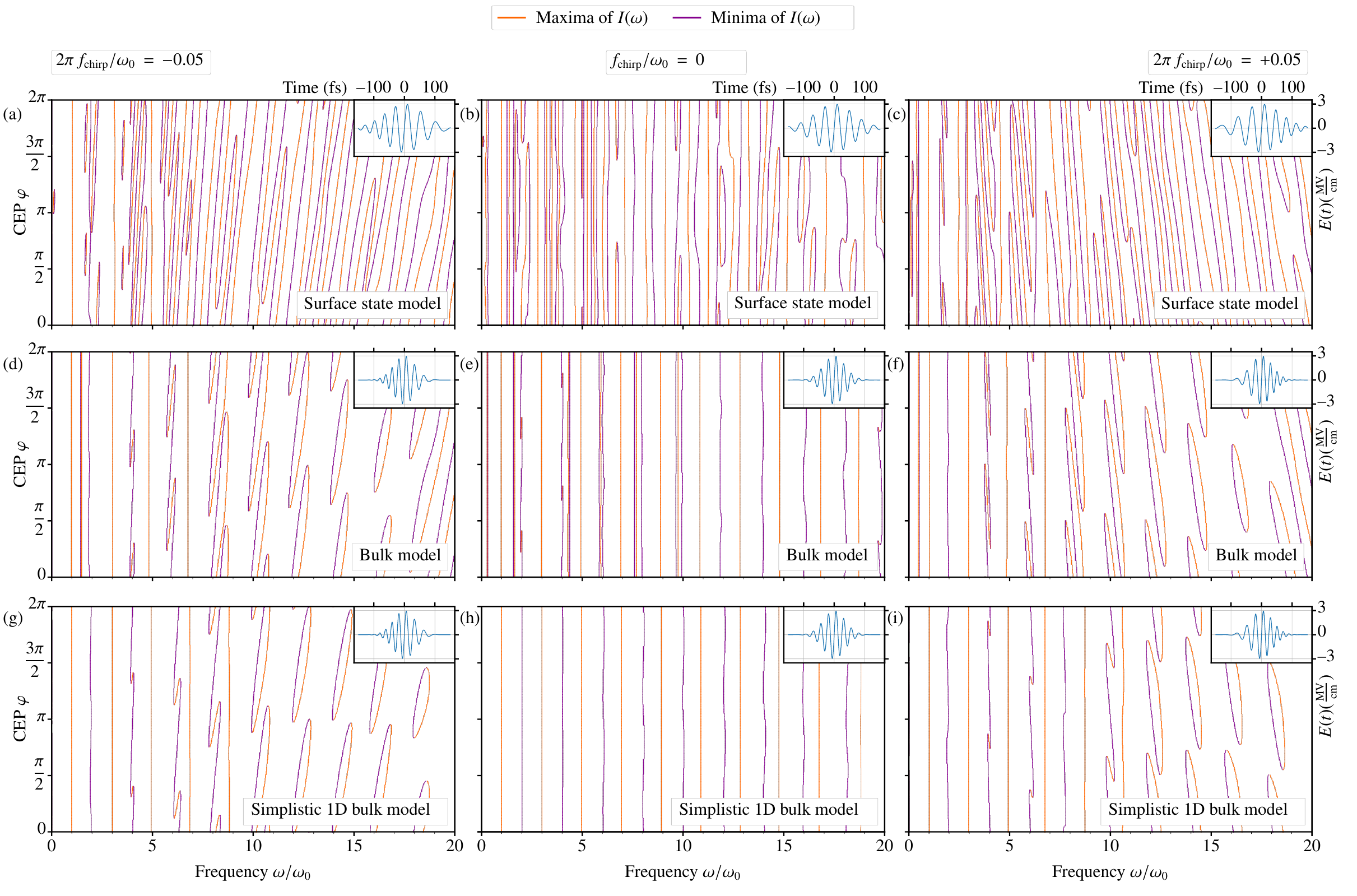}
    \caption{Local extrema of $I(\omega)$ computed from SBE's for the topological surface state (a,b,c) as in Fig.~\ref{f2}, the Bulk Hamiltonian of Bi$_2$Te$_3$ (d,e,f) from Ref.~\cite{Schmid2021} and a one dimensional model with constant
    dipole and symmetric cosine-like bands (g,h,i).
    For the latter, we use a gap of 0.15 eV and a band width of 8.7 eV. The Brillouin zone has a size $2\pi/3\AA$ and dipole is $d\eqt3e\AA$ (e: electric charge).
    We use three different chirp-frequency-ratios, (a,d,g)  $2\pi\,\fchirp / \omega_0\eqt{-}0.05$, (b,e,h) $\fchirp / \omega_0\eqt 0$ and (c,f,i) $2\pi\,\fchirp / \omega_0\eqt{+}0.05$.
    For (a)\,-\,(c), the driving pulse is parametrized from Eq.~\eqref{eq-Efield} with  $E_0\eqt3\,\mathrm{MV/cm}$, $\omega_0\eqt2\pi\cdott25\,\mathrm{THz}$ and $\sigma\eqt90\,\mathrm{fs}$, as already used in Fig.~\ref{f2}.
    For (d)\,-\,(i), the parameters are $E_0\eqt3\,\mathrm{MV/cm}$, $\omega_0\eqt2\pi\cdott40\,\mathrm{THz}$ and $\sigma\eqt50\,\mathrm{fs}$ as in Ref.~\cite{Schmid2021} for simulating HHG from the semiconducting bulk of Bi$_2$Te$_3$. 
    The respective electric fields for CEP $\varphi\eqt0$ are shown in the insets. We employ a dephasing time $T_2\eqt 10$ fs for the surface state model [(a\,-\,c)] and $T_2\eqt 1$ fs in the bulk case [(d\,-\,i)] as in Ref.~\cite{Schmid2021}. 
    Additionally, the band occupations are damped towards the ground state with a damping time~$T_1\eqt10$\,fs [only for (d)\,-\,(i)]  as in Ref.~\cite{Schmid2021}. 
    }
    \label{fig-comparisonbite-tssbulk1d}
\end{figure*}

Furthermore, we report CEP-dependent high-harmonic spectra~$I(\omega)$ for other parameters in the SBE simulations:
$T_2\intext \{1\,\text{fs},\,100\, \mathrm{fs}\}$ in Fig.~\ref{fig-T2comp}, field strength $E_0\intext\{0.1\,\text{MV/cm}, 30\, \mathrm{MV/cm}\}$ in Fig.~\ref{fig-E0comp},  and in addition, semiconducting Hamiltonians in Fig.~\ref{fig-comparisonbite-tssbulk1d}.
%
%We report simulations  in  for two different chirps $\fchirp \eqt \pm 1.25 \, \mathrm{THz}$ .
%
We observe in Figs.~\ref{fig-T2comp} and~\ref{fig-E0comp} that the tilt angle~$d\omega/d\varphi$ increases with increasing harmonic order and that the tilt angle is reversed when changing the sign of the chirp.
Both observations are fully in line with our analytical formula~\eqref{e19}.
We calculate the high-harmonic spectrum from two semiconductor models in Fig.~\ref{fig-comparisonbite-tssbulk1d}\,(d)-(f) and~(g)-(i), in comparison to a model for the topological surface state, Fig.~\ref{fig-comparisonbite-tssbulk1d}\,(a)-(c). 
The model underlying Fig.~\ref{fig-comparisonbite-tssbulk1d}\,(d)-(f) is the bulk semiconductor model for Bi$_2$Te$_3$ from Ref.~\cite{Schmid2021}; the semiconductor model underlying Fig.~\ref{fig-comparisonbite-tssbulk1d}\,(g)-(i) is one-dimensional and has constant dipole and symmetric cosine-like bands.
 We choose a very short dephasing time for the bulk systems, $T_2\eqt1$ fs, and very short damping time of band occupations towards the ground state, $T_1\eqt10$ fs.~\footnote{
Scattering mechanisms are expected to be much more efficient in semiconductors and in the semiconducting bulk of topological insulators (Ref.~\cite{Schmid2021}) compared to topological surface states with spin-momentum locking. }
For semiconductors, we choose the same pulse shape as it has been used for modeling the semiconducting bulk of Bi$_2$Te$_3$ in Ref.~\cite{Schmid2021}.
For all three Hamiltonians, we observe similar tilt angles, for negative chirp to the right [Fig.~\ref{fig-comparisonbite-tssbulk1d}\,(a), (d), (g)], for positive chirp to the left [Fig.~\ref{fig-comparisonbite-tssbulk1d}\,(c), (f), (i)], and for zero chirp almost no tilt [Fig.~\ref{fig-comparisonbite-tssbulk1d}\,(b), (e), (h)], in line with our analytical formula~\eqref{e19}.
\onecolumngrid

\clearpage
\twocolumngrid

\bibliography{Literature}
\bibliographystyle{apsrev4-2.bst}
\end{document}